\documentclass[article,shortnames,nojss]{jss}

\usepackage{amsmath}
\usepackage{amssymb}
\usepackage{amsthm}
\usepackage{soul}

\newcommand{\bTheta}{\boldsymbol{\Theta}}
\newcommand{\bV}{\boldsymbol{V}}
\newcommand{\bU}{\boldsymbol{U}}
\newcommand{\bD}{\boldsymbol{D}}
\newcommand{\bX}{\boldsymbol{X}}
\newcommand{\bY}{\boldsymbol{Y}}

\newcommand{\bZ}{\boldsymbol{Z}}
\newcommand{\tbTheta}{\tilde{\boldsymbol{\Theta}}}
\newcommand{\tbV}{\tilde{\boldsymbol{V}}}
\newcommand{\tbZ}{\tilde{\boldsymbol{Z}}}

\newtheorem{proposition}{Proposition}

\title{\pkg{PReMiuM}: An \proglang{R} Package for Profile Regression Mixture Models using Dirichlet Processes} 
\author{Silvia Liverani$^\ast$\\Imperial College London\\ and\\ MRC Biostatistics Unit\\Cambridge\\$^\ast$ Joint first author\And David I. Hastie$^\ast$\\Imperial College London \\$^\ast$ Joint first author\\ \AND Lamiae Azizi\\MRC Biostatistics Unit\\ Cambridge \And Michail Papathomas\\University of St Andrews\And Sylvia Richardson\\MRC Biostatistics Unit\\Cambridge}

\Plainauthor{Silvia Liverani, David I. Hastie, Lamiae Azizi, Michail Papathomas, Sylvia Richardson} %% comma-separated
\Plaintitle{PReMiuM: An R Package for Profile Regression Mixture Models using Dirichlet Processes} 
\Shorttitle{\pkg{PReMiuM}: An \proglang{R} Package for Profile Regression}

\Abstract{

\pkg{PReMiuM} is a recently developed \proglang{R} package for Bayesian clustering using a Dirichlet process mixture model. This model is an alternative to regression models, non-parametrically linking a response vector to covariate data through cluster membership \citep{MPJ10}. The package allows binary, categorical, count and continuous response, as well as continuous and discrete covariates. Additionally, predictions may be made for the response, and missing values for the covariates are handled. Several samplers and label switching moves are implemented along with diagnostic tools to assess convergence. A number of \proglang{R} functions for post-processing of the output are also provided. In addition to fitting mixtures, it may additionally be of interest to determine which covariates actively drive the mixture components. This is implemented in the package as variable selection.}

\Plainkeywords{Profile regression, Clustering, Dirichlet process mixture model}

\Address{
  Sylvia Richardson\\
  MRC Biostatistics Unit\\
  Cambridge, UK\\
  E-mail: \email{sylvia.richardson@mrc-bsu.cam.ac.uk}\\
}

\begin{document}

\section[Introduction]{Introduction}
\label{sec:intro}
Profile regression is an alternative to regression models when one wishes to make inference beyond main effects for datasets with potentially correlated covariates. In particular, profile regression non-parametrically links a response vector to covariate data through cluster membership \citep{MPJ10}. We have implemented this method in the \proglang{R} \citep{R} package \pkg{PReMiuM}. 

\pkg{PReMiuM} performs Bayesian clustering using a Dirichlet process mixture model and it allows binary, categorical, count and continuous response, as well as continuous and discrete covariates. Moreover, predictions may be made for the response, and missing values for the covariates are handled. Several samplers and label switching moves are implemented along with diagnostic tools to assess convergence. A number of \proglang{R} functions for post-processing of the output are also provided. In addition to fitting mixtures, it may additionally be of interest to determine which covariates actively drive the mixture components. This is implemented in the package as variable selection.

In order to demonstrate the \pkg{PReMiuM} package, it is helpful to present an overview of the Dirichlet Process. We begin this section by re-familiarising the reader with such a process, introducing notation that we shall call upon throughout the paper. Formally, let $(\Omega,\mathcal{F},P)$ be a probability space comprising a state space $\Omega$ with associated $\sigma$-field $\mathcal{F}$ and a probability measure $P$. We say that a probability measure $P$ follows a Dirichlet process with concentration parameter $\alpha$ and base distribution $P_{\Theta_0}$ parametrised by $\Theta_0$, written $P~\sim\mathrm{DP}(\alpha,P_{\Theta_0})$ if \begin{equation}
\label{eqn:dp}
(P(A_1),P(A_2),\ldots,P(A_r))\sim
\mathrm{Dirichlet}(\alpha P_{\Theta_0}(A_1),\alpha
P_{\Theta_0}(A_2),\ldots,\alpha P_{\Theta_0}(A_r)) 
\end{equation}
for all $A_1,A_2,\ldots,A_r\in\mathcal{F}$ such that $A_i\cap
A_j=\emptyset$ for all  $i\neq j$ and $\bigcup_{j=1}^r A_j =\Omega$.

\subsection{The stick-breaking construction}
Although Definition \ref{eqn:dp} is perhaps rather abstract, proof of the existence of such a process has been determined in a variety of ways, using a number of different formulations (\citeauthor{F73}, \citeyear{F73} and \citeauthor{BM73}, \citeyear{BM73}). In this paper we focus on Dirichlet process mixture models (DPMM), based upon the following simplified constructive definition of the Dirichlet process, due to \cite{S94}. If 
\begin{eqnarray}
 \nonumber 
 P & = & \sum_{c=1}^{\infty}\psi_c\delta_{\Theta_c},\\
 \nonumber 
 \Theta_c &\sim& P_{\Theta_0} \;\;\textrm{i.i.d. for }c\in\mathbb{Z}^{+},\\
 \label{eqn:stick}
 \psi_c & = & V_c\prod_{l<c}(1-V_l)\;\;\textrm{for
}c\in\mathbb{Z}^{+}\setminus\{1\},\\
 \nonumber 
 \psi_1 & = & V_1,\;\;\textrm{ and}\\
 \nonumber
 V_c & \sim & \mathrm{Beta}(1,\alpha)\;\;\textrm{i.i.d. for }c\in\mathbb{Z}^{+}, 
\end{eqnarray}
where $\delta_x$ denotes the Dirac delta function concentrated at $x$ and $\Theta_c$ is independent of $V_c$ for $c \in \mathbb{Z}^{+}$, then $P~\sim\mathrm{DP}(\alpha,P_{\Theta_0})$. This formulation for $\bV$ and $\boldsymbol{\psi}$ is known as a \emph{stick-breaking} distribution. Importantly, the distribution $P$ is discrete, because draws $\tilde{\Theta}_1,\tilde{\Theta}_2,\ldots$ from $P$ can only take the values in the set $\{\Theta_c:c\in\mathbb{Z}^{+}\}$.

As noted by many authors (for example \citeauthor{IJ01}, \citeyear{IJ01} and \citeauthor{KGW11}, \citeyear{KGW11}) it is possible to extend the above formulation to more general stick-breaking formulations, for example allowing $V_c\sim\mathrm{Beta}(a_c,b_c)$ independently, resulting in a generalised Dirichlet process, such as the two parameter Poisson-Dirichlet process \citep{PY97}. %Another example is provided by \cite{CD09} and \cite{RD11} where the Beta distibutions are replaced by a Probit model, which may be allowed to depend on observed data, for the purpose of capturing dependence such as a spatial structure. 
The methods and results that we propose within this paper hold for such generalised processes, but at present the package is only coded to implement the Dirichlet process in Equation~\ref{eqn:stick} and the Poisson-Dirichlet process with $V_c\sim\mathrm{Beta}(1-d,\alpha-cd)$ where $d \in [0,1)$ and $\alpha > -d$. For $d=0$ the Dirichlet process is a special case of the Poisson-Dirichlet process. 

Typically, because of the complexity of the models based on the stick-breaking construction, inference is made in a Bayesian framework using Markov chain Monte Carlo (MCMC) methods. Until recently, a perceived difficulty in making inference about this model was the infinite number of parameters within the stick breaking construction. Historically, this obstacle has resulted in the use of algorithms that either explore marginal spaces where some parameters are integrated out or use truncated approximations to the full Dirichlet process mixture model, see for example \cite{N00} and \cite{IJ01}.

More recently, two alternative innovative approaches to sampling the full DPMM have been proposed. The first, introduced by \cite{W07}, uses a novel slice sampling approach, resulting in full conditionals that may be explored by the use of a Gibbs sampler. The slice sampling method updates the cluster allocations jointly as opposed to the marginal methods which requires as many Gibbs steps to update as the number of observations to cluster. The difficulty of the proposed approach is the introduction of constraints that complicate the updates of the mixture component weights, leading to potential mixing issues. To overcome this, \cite{KGW11} generalise this sampler, adding further auxilliary variables, and report good convergence results, although the authors note that the algorithm is sensitive to these additional parameters. The second distinct MCMC sampling approach was proposed in parallel by \cite{PR08}. The proposed sampler again uses a Gibbs sampling approach, but is based upon an idea termed \emph{retrospective sampling}, allowing a dynamic approach to the determination of the number of components (and their parameters) that adapts as the sampler progresses. The cost of this approach is an ingenious but complex Metropolis-within-Gibbs step, to determine cluster membership.

Despite the apparent differences between the two strategies, \cite{P08} noted that the two algorithms can be effectively combined to yield an algorithm that improves either of the originals. The resulting sampler was implemented and presented by \cite{YPR11}, and a similar version was presented by \cite{D09} for DPMM. The current sampler presented in this paper is our interpretation of these ideas, implemented as an \proglang{R} package. This package, called \pkg{PReMiuM}, is based upon efficient underlying \proglang{C++} code for general DPMM sampling and it is available on CRAN.

The aims behind the Product Partition Model (PPMx) in \cite{muller11} and \cite{quintana13} are very similar to ours. Furthermore, both sets of work adopt flexible Bayesian partition models based on the Dirichlet process (DP), although the PPMx approach is adaptable to formulations other than the DP. However, there are significant differences in how the two models are built. For example, the PPMx model is built by considering the likelihood of the partition given the covariates and variable selection parameters, using similarity functions. In contrast, we consider the likelihood of the covariates given the partition and variable selection parameters. 
%There are other differences too; for example, the authors consider the logit of the binary cluster specific selection switches and impose an additional level in the hierarchy using Normal densities and associated hyper-parameters. A normalization step is then required. In contrast, we impose an additional level in the hierarchy considering Bernoulli distributions for the binary switches, without the requirement of a normalization step. 
The two modelling approaches offer different options for defining the dependence structure between the quantities of interest, and we would argue that it is a matter of personal preference which one should be adopted. 

In Section \ref{sec:dpmm} we describe formally the Dirichlet process mixture model implemented in \pkg{PReMiuM} and the blocked MCMC sampler used. In Section \ref{sec:models} we discuss profile regression and its link with the response and covariate models included in the package while in Section \ref{section:preds} we discuss how predictions are computed. In Section \ref{sec:postprocessing} we give an overview of the postprocessing tools available in \pkg{PReMiuM} to learn from the rich output produced by our Bayesian model and in Section \ref{sec:mixing} we discuss diagnostic tools that we propose to investigate the convergence of the MCMC. Finally, in Section \ref{sec:software} we give a brief overview of the structure of the code and show examples of its use in Section \ref{sec:example}. We also give an indication of run times.

\section{Sampling the Dirichlet process mixture model}
\label{sec:dpmm}
\subsection{Definition and properties}
\label{subsec:dpmmdefn}
Perhaps the most common application of the Dirichlet process is in clustering data, where it can be used as the prior distribution for the parameters of an infinite mixture model. Consider again the stick breaking construction in Equation~\ref{eqn:stick}. For the Dirichlet process mixture model (DPMM), the (possibly multivariate) observed data $\bD=(D_1,D_2,\ldots,D_n)$ follow an infinite mixture distribution, where component $c$ of the mixture is a parametric density of the form $f_c(\cdot)=f(\cdot|\Theta_c,\Lambda)$ parametrised by some component specific parameter $\Theta_c$ and some global parameter $\Lambda$. Defining (latent) parameters $\tilde{\Theta}_1,\tilde{\Theta}_2,\ldots,\tilde{\Theta}_n$ as draws from a probability distribution $P$ following a Dirichlet process $DP(\alpha,P_{\Theta_0})$ and again denoting the Dirac delta function by $\delta$, this system can be written as
\begin{eqnarray}
\nonumber
  D_i|\tilde{\Theta}_i,\Lambda &\sim
&f(D_i|\tilde{\Theta}_i,\Lambda)\;\;\textrm{for
}i=1,2,\ldots,n,\\
 \nonumber
 \tilde{\Theta}_i & \sim & \sum_{c=1}^{\infty}\psi_c\delta_{\Theta_c}
\;\;\textrm{for }i=1,2,\ldots,n,\\
 \nonumber 
 \Theta_c &\sim& P_{\Theta_0} \;\;\textrm{i.i.d. for }c\in\mathbb{Z}^{+},\\
 \label{eqn:dpmm1}
 \psi_c & = & V_c\prod_{l<c}(1-V_l)\;\;\textrm{for
}c\in\mathbb{Z}^{+}\setminus\{1\},\\ 
 \nonumber 
 \psi_1 & = & V_1,\;\;\textrm{and}\\ 
 \nonumber
 V_c & \sim & \mathrm{Beta}(1,\alpha)\;\;\textrm{i.i.d. for }c\in\mathbb{Z}^{+}.
\end{eqnarray}
with $\Theta_c$ independent of $V_c$ for $c \in \mathbb{Z}^{+}$.

When making inference using mixture models (either finite or infinite) it is common practice to introduce a vector of latent allocation variables $\bZ$. Such variables enable us to explicity characterise the clustering and additionally facilitate the design of MCMC samplers. Adopting this approach and writing $\boldsymbol{\psi}=(\psi_1,\psi_2,\ldots)$ and $\bTheta=(\Theta_1,\Theta_2,\ldots)$, we re-write Equation~\ref{eqn:dpmm1} as 
\begin{eqnarray}
\nonumber
  D_i|\bZ,\bTheta,\Lambda &\sim
&f(D_i|\Theta_{Z_i},\Lambda)\;\;\textrm{for }i=1,2,\ldots,n,\\
 \nonumber 
 \Theta_c &\sim& P_{\Theta_0} \;\;\textrm{i.i.d. for }c\in\mathbb{Z}^{+},\\
 \nonumber
 \mathbb{P}(Z_i=c|\boldsymbol{\psi}) & = & \psi_c \;\;\textrm{for
}c\in\mathbb{Z}^{+},\;i=1,2,\ldots,n,\\
 \label{eqn:dpmm2}
 \psi_c & = & V_c\prod_{l<c}(1-V_l)\;\;\textrm{for
}c\in\mathbb{Z}^{+}\setminus\{1\},\\ 
 \nonumber 
 \psi_1 & = & V_1,\;\;\textrm{and}\\ 
 \nonumber
 V_c & \sim & \mathrm{Beta}(1,\alpha)\;\;\textrm{i.i.d. for }c\in\mathbb{Z}^{+}.
\end{eqnarray}
with $\Theta_c$ independent of $V_c$ for $c \in \mathbb{Z}^{+}$.

The likelihood of $D_i$ associated with the DPMM is simply the first line of Equation~\ref{eqn:dpmm2}. Integrating out the latent variable $Z_i$ we obtain the more recognisable mixture likelihood
\begin{eqnarray*}
%\nonumber
p(D_i|\boldsymbol{\psi},\bTheta,\Lambda)&
= &
\sum_{c=1}^{\infty}p(D_i|Z_i=c,\bTheta,
\Lambda)p(Z_i=c|\boldsymbol{\psi})\\
%\nonumber
& = &\sum_{c=1}^{\infty} \psi_cf(D_i|\Theta_c,\Lambda).
\end{eqnarray*}

The remainder of Equation~\ref{eqn:dpmm2} provides the prior specification for the DPMM, allowing us to write the joint posterior distribution as
\begin{eqnarray}
\nonumber
 p(\bZ,\bTheta,\Lambda,\bV,
\alpha|\bD)&\propto&p(\bD|\bZ,\bTheta,\Lambda)p(\bZ,\bV,\alpha,\bTheta,\Lambda|\Theta_0)\\
\nonumber
& \propto & \prod_{i=1}^n \left\{f(D_i|\Theta_{Z_i},\Lambda)
p(Z_i|\bV)\right\}\prod_{c=1}^{\infty}
\left\{p(V_c|\alpha)p(\Theta_c|\Theta_0)\right\}p(\Lambda)p(\alpha)\\
\label{eqn:dpmmpost}
& \propto & \prod_{i=1}^n \left\{f(D_i|\Theta_{Z_i},\Lambda)
\left[V_{Z_i}\prod_{l<Z_i}(1-V_l)\right]\right\}\\
\nonumber
& & \;\;\;\;\;\;\;\;\;\;\;\;\;\;\times\prod_{c=1}^{\infty}
\left\{\alpha(1-V_c)^{\alpha-1}p(\Theta_c|\Theta_0)\right\}
p(\Lambda)p(\alpha).
\end{eqnarray}
Of course, additional layers of hierarchy could easily be introduced, for example through hyper-priors for $\Theta_0$. 

\subsection{The MCMC sampler}
\label{subsec:blocksampler}
A common approach to MCMC sampling from the DPMM is to integrate out $\bV$ and use a Gibbs sampler on the resulting space. Such samplers are commonly referred to as \emph{P\'{o}lya urn} samplers, since they are motivated by the P\'{o}lya urn representation of a Dirichlet process introduced by \cite{BM73}. \cite{IJ01} provide a review of this approach and demonstrate that conditionals of the type $p(Z_i|\bZ_{-i})$ can be derived, where $\bZ_{-i}=(Z_1,\ldots,Z_{i-1},Z_{i+1},Z_n)$. Many other authors have focused on developing alternative samplers of this nature, including \cite{N00} and \cite{G10}. % and \cite{FBT11}. 
However, as noted by many authors, samplers of this nature, where the allocation of a single observation is conditional on the allocations of all other observations, can often suffer from poor mixing. This motivates the need for an alternative class of samplers that sample from the full model in Equation~\ref{eqn:dpmm2}. 

In the full model, the posterior conditionals for the allocation variables $\bZ$ depend upon an infinite number of variables $\bV$ and $\bTheta$. One way to bypass this complication is to truncate the definition in Equation~\ref{eqn:dpmm2} to mixtures with $C$ components (of which potentially only a subset will be non-empty). \cite{IJ01} demonstrate that under this model the conditional distributions are standard distributions which can be easily sampled from. This is one of the three samplers implemented in our R package, and we refer to it as \textit{truncated}. For the truncated sampler we have also implemented the more general Poisson Dirichlet stick-breaking formulation \citep{PY97}, constructed by allowing $V_c\sim\mathrm{Beta}(1-d,\alpha-cd)$ where $d \in [0,1)$ and $\alpha > -d$ in Equation~\ref{eqn:stick}. For this model $\alpha$ and $d$ are fixed parameters. 

Although the approach of \cite{IJ01} resolves the challenges of sampling from the full model, if $C$ is not chosen to be sufficiently large, then the posterior may be quite different on the truncated model space compared to the full model space. The authors provide some guidance for choice of $C$, but more recent work by \cite{W07} and \cite{PR08} demonstrate techniques that alleviate the need for such a truncation, whilst retaining many of the sampling properties of the full conditionals.

The first step is to borrow the idea of \cite{W07} and introduce auxiliary variables $\bU=(U_1,U_2,\ldots,U_n)$ such that the joint posterior can be re-written as 
\begin{eqnarray}
 p(\bU,\bZ,\bTheta,\Lambda,\bV ,
\alpha|\bD)&\propto& \prod_{i=1}^n
\left\{f(D_i|\Theta_{Z_i},\Lambda)\boldsymbol{1}_{\{U_i<\psi_{Z_i}
\}}\right\}\\
\nonumber
& & \;\;\;\;\;\;\;\;\;\;\;\;\;\;\times\prod_{c=1}^{\infty}
\left\{\alpha(1-V_c)^{\alpha-1}p(\Theta_c|\Theta_0)\right\}
p(\Lambda)p(\alpha)\\
\label{eqn:dpmmpostu}
&\propto& \prod_{i=1}^n
\left\{f(D_i|\Theta_{Z_i},\Lambda)\boldsymbol{1}_{\{U_i<V_{Z_i}\prod_{
l<Z_i} (1-V_l)\}}\right\}\\
\nonumber
& & \;\;\;\;\;\;\;\;\;\;\;\;\;\;\times\prod_{c=1}^{\infty}
\left\{\alpha(1-V_c)^{\alpha-1}p(\Theta_c|\Theta_0)\right\}
p(\Lambda)p(\alpha).
\end{eqnarray}
Here, $\boldsymbol{1}_A$, is the function that takes the value 1 over the set $A$ and 0 elsewhere. By combining this auxiliary variable approach with the notion of retrospective sampling (i.e., adopting a just-in-time approach to sampling empty mixture component parameters, as introduced in \citeauthor{PR08}, \citeyear{PR08}), it is possible to construct an efficient Gibbs sampler for sampling from the joint posterior in Equation~\ref{eqn:dpmmpostu}. Integrating $\bU$ out of Equation~\ref{eqn:dpmmpostu} with respect to the Lebesgue measure yields the DPMM posterior distribution given in Equation~\ref{eqn:dpmmpost} meaning that marginalising samples of the joint distribution over $\bU$ results in samples from the desired distribution. 

\cite{KGW11} extend the idea of \cite{W07} to a general class of slice samplers by writing 
\begin{eqnarray}
 p(\bU,\bZ,\bTheta,\Lambda,\bV ,
\alpha|\bD)&\propto& \prod_{i=1}^n
\left\{f(D_i|\Theta_{Z_i},\Lambda)\boldsymbol{1}_{\{U_i<\xi_{Z_i}
\}}\xi_{Z_i}^{-1}\psi_{Z_i}\right\}\\
\nonumber
& & \;\;\;\;\;\;\;\;\;\;\;\;\;\;\times\prod_{c=1}^{\infty}
\left\{\alpha(1-V_c)^{\alpha-1}p(\Theta_c|\Theta_0)\right\}
p(\Lambda)p(\alpha)
\label{eqn:dpmmpostupostKalli}
\end{eqnarray}
where $\xi_1,\xi_2, \ldots$ is any positive sequence. 

When $\xi_i=\psi_i$, this corresponds to the efficient Gibbs sampler proposed by \cite{P08} in the context of DPMM using parameter blocking. This is the second of the three samplers implemented in our R package, and we refer to it as \textit{slice dependent}, in accordance with \cite{KGW11}. 

For the last of the three samplers implemented in our R package, that we refer to as \textit{slice independent} in accordance with \cite{KGW11}, we set $\xi_i=(\kappa-1) \kappa^{i-1}$ with $\kappa = 0.8$ as proposed by \cite{KGW11}.

Most importantly, these slice samplers permit the introduction of label switching moves, without which it is very difficult to obtain sufficient mixing. We discuss this in detail in Section \ref{sec:mixing}. 

We continue by defining some new notation which is required to present our slice samplers. First, given the allocation variables $\bZ$, define
\[
 Z^{\star} = \max_{1\leq i \leq n} Z_i.
\]
Similarly, given the auxiliary variables $\bU$ and the vector $\bV$, define 
\[
 U^{\star} = \min_{1\leq i \leq n} U_i.
\]
and
\begin{eqnarray}
\label{eqn:cstar} 
C^{\star} & = & \min\left\{c\in\mathbb{Z}^+: \sum_{l=1}^c \psi_l > 1 - U^{\star}
\right\}\\
\nonumber
& = & \min\left\{c\in\mathbb{Z}^+: \sum_{l=1}^c
\left[V_l\prod_{r<l}(1-V_r)\right] > 1 - U^{\star}\right\}.                 
\end{eqnarray}
It is important to emphasise that these values potentially change at each sweep of the sampler, as the underlying variables change, although for simplicity of exposition we have omitted explicitly labelling the parameters with the sweep. The purpose of the variable $C^{\star}$ is to provide an upper limit on which mixture components need updating at each sweep. Specifically, although there are infinitely many component parameters in the model, since $P(Z_i=c|U_i>\psi_c)=0$, we need only concentrate our updating efforts on those components $c$ for which $\psi_c>U_i$ for some $i=1,2,\ldots,n$. By defining $C^{\star}$ as in Equation~\ref{eqn:cstar} it can be shown (see Appendix \ref{appendix:cstar}) that $\psi_c<U_i$ for all $c>C^{\star}$ and all $i=1,2,\ldots,n$. Assuming that the Markov chain is initialised accordingly and is updated using the correct conditionals, it is possible to show $Z^{\star} \leq C^{\star} < \infty$ almost surely (details provided in Appendix \ref{appendix:cstar}). 

With these definitions in place we make use of the following sets and vectors (which again will change at each sweep)
\begin{eqnarray}
 \nonumber & A = \{c\in\mathbb{Z}^{+}:c\leq Z^{\star}\},\;\;P =
\{c\in\mathbb{Z}^{+}:Z^{\star}<c \leq C^{\star}\},\;\;I =
\{c\in\mathbb{Z}^{+}:c> C^{\star}\} & \\
 \nonumber
 & \bV^{A}=(V_1,V_2,\ldots,V_{Z^{\star}}),\;\;\;
 \bTheta^{A}=(\Theta_1,\Theta_2,\ldots,
\Theta_{Z^{\star}})&\\
 \nonumber
 &\bV^{P}=(V_{Z^{\star}+1},V_{Z^{\star}+2},\ldots,V_{C^{\star}}
),\;\;\;\bTheta^{P}=(\Theta_{Z^{\star}+1},
\Theta_ {Z^{\star}+2},\ldots,\Theta_{C^{\star} })&\\
 \nonumber
 &\bV^{I}=(V_{C^{\star}+1},V_{C^{\star}+2},\ldots),\;\;\;
\bTheta^{I}=(\Theta_{C^{\star}+1},\Theta_{C^{\star}+2},\ldots).&
\end{eqnarray}
Here the $A$, $P$ and $I$ are disjoint sets (updated at every sweep of the MCMC algorithm) that partition $\mathbb{Z}^{+}$, with names chosen to denote \emph{Active}, \emph{Potential} and \emph{Inactive} components respectively. It is possible that $P=\emptyset$. By definition, all observations are allocated to a mixture component labelled by one of the indices in $A$. Components labelled with an index in $P$ or $I$ are necessarily empty (i.e., have no observations allocated to them), the difference being that at the next update of the allocation variables, components with labels in $P$ may potentially become non-empty. 

The blocked infinite DPMM algorithm can now be defined using the following blocked Gibbs updates to sample from the relevant conditionals.

\subsubsection{DPMM algorithm}
Suppose we are at sweep $t$ of the sampler. Update as follows:
\begin{description}
\item[\emph{Step A.}]{Compute $Z^{\star}$ and the set $A$}.  
\item[\emph{Step B.}]{Sample
$(\bV^A_{t+1},\bTheta^A_{t+1}, \tbZ,\bU_{t+1})\sim 
p(\bV^A, \bTheta^A,\bZ,\bU |\bV^P_t,
\bV^I_t,\bTheta^P_t,\bTheta^I_t,\alpha_t,\Lambda_t,\Theta_0,\bD).$
\begin{description}
 \item[\emph{B.1}]{$\tbV^A\sim p(\bV^A|\bZ_t,\alpha_t)$}
 \item[\emph{B.2}]{$\tbTheta^A\sim p(\bTheta^A|\bZ_t,\Lambda_t,\Theta_0,\bD)$}
 \item[\emph{B.3}]{$(\bV^A_{t+1},\bTheta^A_{t+1},\tbZ)\sim
p(\bV^A,\bTheta^A,\bZ|
\bV^P_t,\bTheta^P_t,\alpha_t,\Lambda_t,\Theta_0,\bD)$}
\item[\emph{B.4}]{$\bU_{t+1}\sim p(\bU|\bV^A_{t+1},\tbZ)$}
\end{description}}
\item[\emph{Step C.}]{Compute $U^{\star}$. Recompute $Z^{\star}$ and the set
$A$.}
\item[\emph{Step D.}]{Sample
$(\alpha_{t+1},\bV^P_{t+1},\bV^I_{t+1})\sim
p(\alpha,\bV^P,\bV^I|\bTheta^P_{t+1},\bV^A_{t+1},\bTheta^A_{t+1},\bTheta^I_t,\bU_{t+1},\tbZ,
\Lambda_t,\Theta_0,\bD)$, computing $C^{\star}$ and the set $P$ in the process.
\begin{description}
 \item[\emph{D.1}]{$\alpha_{t+1}\sim p(\alpha|\bV^A_{t+1},\tbZ)$}
 \item[\emph{D.2}]{$\bV^P_{t+1}\sim p(\bV^P|\alpha_{t+1},\bU_{t+1},\tbZ)$}
\end{description}}
\item[\emph{Step E.}]{Sample 
$(\bTheta^P_{t+1},\bTheta^I_{t+1})\sim p(\bTheta^P,\bTheta^I|\bV^A_{t+1},\bV^P_
{t+1},\bV^I_{t+1},\bTheta^A_{t+1},\bU_{t+1},\tbZ,\alpha_{t+1},
\Lambda_t,\bTheta_0,\bD).$
\begin{description}
 \item[\emph{E.1}]{$\bTheta^P_{t+1}\sim p(\bTheta^P|\Theta_0)$}
\end{description}
}
\item[\emph{Step F.}]{Sample 
$\Lambda_{t+1}\sim p(\Lambda|\bV^A_{t+1},\bV^P_
{t+1},\bV^I_{t+1},\bTheta^A_{t+1},\bTheta^P_{t+1},\bTheta^I_{t+1},\bU_{t+1},\tbZ,\alpha_{
t+1},\Theta_0,\bD).$
\begin{description}
 \item[\emph{F.1}]{$\Lambda\sim p(\Lambda|\bTheta^A_{t+1},\tbZ,\bD)$}
\end{description}}
\item[\emph{Step G.}]{Sample 
$\bZ_{t+1}\sim
p(\bZ|\bV^A_{t+1},\bV^P_{t+1},\bV^I_{t+1},\bTheta^A_{t+1},\bTheta^P_{t+1},
\bTheta^I_{t+1},\bU_{t+1},\alpha_{t+1},
\Lambda_{t+1},\Theta_0,\bD).$
\begin{description}
 \item[\emph{G.1}]{$\bZ_{t+1}\sim
p(\bZ|\bV^A_{t+1},\bV^P_{t+1},\bTheta^A_{t+1},\bTheta^P_{t+1},\bU_{t+1},
\Lambda_{t+1},\bD)$}
\end{description}}
\end{description}

While this algorithm is somewhat generic, the blocking strategy is clearly highlighted. Further details explaining each of the steps are provided in Appendix \ref{appendix:algo1}. The key idea is that by doing joint updates, we can marginalise out an infinite number of variables when necessary, to ensure that we are always sampling from conditional distributions that depend only upon a finite number of parameters. In particular, after marginalisation, the parameters corresponding to the inactive set $I$ do not contribute to the conditional distributions of the other parameters, so we do not actually need to sample their values. Since these parameters have no contribution to the likelihood, if values are subsequently required they can simply be sampled from the prior retrospectively as necessary. Although the sampler is written as a blocked Gibbs sampler, where it is not possible to sample directly from full conditionals (for example in the update of $\bTheta$, depending upon the choices of $f$ and $P_{\Theta_0}$) Metropolis-within-Gibbs steps are applied. Depending on the application, the Gibbs updates specified above may comprise several  different Gibbs or Metropolis-within-Gibbs steps (for example updating $\bTheta$ and $\Lambda$. Typically, where Metropolis-Hastings updates are required we advocate adopting an adaptive Metropolis-Hastings approach: see \cite{AT08} for a review.

\section{Example models}
\label{sec:models}
The general sampler of the previous section is applicable for many specific models, depending on the choices of $f$ and $P_{\Theta_0}$. In this section we provide further details for some of the models that are implemented within our software. We detail the prior choices that are made within our implementation, but, of course, alternative priors could be chosen.
\subsection{Gaussian mixtures}
\label{subsec:gaussmix}
Perhaps the most common model to be implemented under the DPMM framework is the Gaussian mixture model, where $\bD=\bX$ for some covariate data $\bX$, and $\bX$ assumes a mixture of Gaussian distributions. Under this setting for each cluster $c$, the cluster specific parameters are given by $\Theta_c=(\mu_c,\Sigma_c)$, where $\mu_c$ is a mean vector and $\Sigma_c$ is a covariance matrix. There are no additional global parameters $\Lambda$. Under this setting 
\begin{equation}
\label{eqn:gaussmix}
p(X_i|Z_i,\Theta_{Z_i},\Lambda)=f(X_i|\mu_{Z_i},\Sigma_{Z_i})=(2\pi)^{-\frac{J}{2}}|\Sigma_{Z_i}|^{-\frac{1}{2}}\exp\left\{-\frac{1}{2}(X_i-\mu_{Z_i})^\top \Sigma_{Z_i}^{-1}(X_i-\mu_{Z_i})\right\}.
\end{equation}

By choosing $\mu_c\sim\mathrm{Normal}(\mu_0,\Sigma_0)$ and $\Sigma_c\sim\mathrm{InvWishart}(R_0,\kappa_0)$ (for each $c$) for our prior model $P_{\Theta_0}$  we have a conjugate model, permitting Gibbs updates for the parameters $\boldsymbol{\mu}^A$ and $\boldsymbol{\Sigma}^A$ associated with the active clusters, and also those ($\boldsymbol{\mu}^P$ and $\boldsymbol{\Sigma}^P$) associated with the potential clusters. The choice of values for the hyperparameters $\Theta_0=(\mu_0,\Theta_0,R_0,\kappa_0)$ is discussed further in Section \ref{sec:software}. 

\subsection{Discrete mixtures}
\label{subsec:discretemix}
Clearly the DPMM model applies to mixtures other than the Gaussian one. Consider for example the case where for each individual $i$, $D_i=X_i$ is a vector of $J$ locally independent discrete categorical random variables, where the number of categories for covariate $j=1,2,\ldots,J$ is $K_j$. Then we can write $\Theta_c=\Phi_c=(\Phi_{c,1},\Phi_{c,2}\ldots,\Phi_{c,J})$ with $\Phi_{c,j} = (\phi_{c,j,1},\phi_{c,j,2},\ldots,\phi_{c,j,K_j})$ and
\begin{equation}
\label{eqn:discretemix}
   p(D_i|Z_i,\Theta_{Z_i},\Lambda)=f(D_i|\Phi_{Z_i})=\prod_{j=1}^J\phi_{{Z_i},j,X_{i,j}}. 
\end{equation}
Again, there are no global parameters $\Lambda$. 

Letting $\Theta_0=a=(a_1,a_2,\ldots,a_J)$, where for $j=1,2,\ldots,J,\;$ $a_j=(a_{j,1},a_{j,2},\ldots,a_{j,K_j})$ and adopting conjugate Dirichlet priors $\Phi_{c,j}\sim\mathrm{Dirichlet}(a_j)$, each $\Phi_c$ can be updated directly using Gibbs updates. For full details of the posterior conditional distribution see \cite{MPJ10}.  

\subsection{Mixed mixtures}
\label{subsec:mixedmix}
An alternative model is given by a mixture of some continuous and discrete random variables. Following the notation used above for Gaussian and discrete mixtures, for $J_1$ continuous random variables and $J_2$ discrete random variables, 
\begin{equation}
   p(D_i|Z_i,\Theta_{Z_i},\Lambda)=p(D_i^{1}|\mu_{Z_i},\Sigma_{Z_i})p(D_i^{2}|\Phi_{Z_i})
\end{equation}
where $D_i^1$ is the subset of the continuous random variables in $D_i$ and $D_i^2$ is the subset of the categorical random variables in $D_i$. Note that we are assuming independence between continuous and categorical data conditional on the cluster allocations.

\subsection{Profile regression}\label{section:profile}
Recently, interest has grown in using DPMM as an alternative to regression models, non-parametrically linking a response vector $\bY$ to covariate data $\bX$ through cluster membership. This idea has been pioneered by several authors including \cite{DHS08}, \cite{BD09}, \cite{MPJ10}, \cite{PMR11}, and \cite{MSM11}. Our presentation is most similar to the latter three of these articles which refer to this idea as ``profile regression''.

For the case of either Gaussian or discrete mixtures, as described above, our implementation permits the joint modelling of a response vector, for various response models which we present below. Formally, the data $\bD=(\bY,\bX)$ are now extended to contain response data $Y_i$ and covariate data $X_i$ for each individual $i$, where the contribution of the covariate data to the response may be cluster dependent. There is also the possibility to include additional fixed effects $W_i$ for each individual, which are constrained to only have a global (i.e., non-cluster specific) effect on the response $Y_i$. 

The data $D_i$ are then jointly modelled as the product of a response model and a covariate model, to give the following likelihood:
\[
   p(D_i|Z_i,\bTheta,\Lambda,W_i)=f_Y(y_i|\Theta_{Z_i},\Lambda,W_i)f_X(x_i|\Theta_{Z_i},\Lambda).
\]

The covariate likelihood $f_X$ is of either of the forms presented in Sections \ref{subsec:gaussmix} or \ref{subsec:discretemix}. The likelihood $f_Y$ depends upon the choice of response model.

\subsubsection{Binary response}
Adopting a binary response model, each parameter vector $\bTheta_c$ is extended to include an additional parameter $\theta_c$. We also introduce the global parameter vector $\Lambda=\beta$, of the same length $L$ as the fixed effects vector $W_i$, to capture the contribution of these effects. Then, $f_Y(y_i|\bTheta_{Z_i},\Lambda,W_i)=p(Y_i=1|\theta_{Z_i},\beta,W_i)$ is given by
\[
   \mathrm{logit}\{p(Y_i=1|\theta_{Z_i},\beta,W_i)\} :=\lambda_i = \theta_{Z_i} + \beta^\top W_i.
\]
For each cluster $c$, we adopt a $t$ location-scale distribution for $\theta_c$, with hyperparameters $\mu_{\theta}$ and $\sigma_{\theta}$ with 7 degrees of freedom, as discussed by \cite{MPJ10}. Similarly, for each fixed effect $l$, we adopt the same prior for $\beta_l$, but with hyperparameters $\mu_{\beta}$ and $\sigma_{\beta}$. 

The components of $\Theta_c$ corresponding to the covariate model for $\bX$ retain the possibility of being updated according to Gibbs samples. However, since conjugacy is not achieved with our prior choice, updating $\theta_c$ for each cluster (and $\beta_l$ for each fixed effect $l$) requires a Metropolis-within-Gibbs sample. In our implementation we propose the use of adaptive random-walk-Metropolis moves.

\subsubsection{Categorical response}
The categorical response model that we use is a simple extension of the binary response model of the previous section. In particular, each parameter vector $\bTheta_c$ additionally contains an extra parameter vector $\theta_c=(\theta_{c,1},\theta_{c,2},\dots,\theta_{c,R-1})$ of length $R-1$, where $R$ is the number of possible categories represented in the response data $\bY$. Treatment of fixed effects is also extended, so that for each response category $r=1,2,\ldots,R-1$, there is a vector $\beta_{r}=(\beta_{r,1},\beta_{r,2},\ldots,\beta_{r,L})$, where $\beta_{r,l}$ is the coefficient for each fixed effect $l$ ($l=1,2,\ldots,L$). This gives $f_Y(y_i|\bTheta_{Z_i},\Lambda,W_i)=p(Y_i=r|\theta_{{Z_i},r},\beta,W_i)$ as
\[
   \mathrm{logit}\{p(Y_i=r|\theta_{Z_i},\beta,W_i)\} = \theta_{{Z_i},r} + \beta_r^\top W_i, \;\;\;\textrm{for } r=1,2,\dots,R-1
\]
and $p(Y_i=0|\theta_{Z_i},\beta,W_i)=1-\sum_{r=1}^{R-1}p(Y_i=r|\theta_{Z_i},\beta,W_i)$.

In our sampler we use the same priors for each $\theta_{c,r}$ as for $\theta_c$  and $\beta_{r,l}$ as for $\beta_l$ in the binary case, with the resulting observation about Metropolis-within-Gibbs updates remaining true. Note that $\theta_{c,r}$ and $\beta_{c,r}$ are independent across $r$.

\subsubsection{Count response modelled as Binomial}
By providing a number of trials $T_i$ associated with each individual $i$ (in this model an ``individual'' might correspond to an area or ``experiment'') we can extend the binary response model to a Binomial response model. In particular,
\[
f_Y(y_i|\bTheta_{Z_i},\Lambda,W_i)=p(Y_i|\theta_{Z_i},\beta,W_i)={T_i\choose Y_i}p_i^{Y_i}(1-p_i)^{T_i-Y_i},
\]
where
\[
   \mathrm{logit}\{p_i\} := \lambda_i = \theta_{Z_i} + \beta^\top W_i.
\]
Priors used are identical to the binary case. \cite{MSM11} provide an example where this model is employed.

\subsubsection{Count response modelled as Poisson}
For count-type response data, an alternative to the Binomial model is the Poisson model. Under this model, each individual $i$ is associated with an expected offset $E_i$, and the response is then modelled as
\[
f_Y(y_i|\bTheta_{Z_i},\Lambda,W_i)=p(Y_i|\theta_{Z_i},\beta,W_i)=\frac{\mu_i^{Y_i}}{Y_i!}\exp\{-\mu_i\},
\]
where
\[ 
\mu_i=E_i \exp\{\lambda_i\},\;\;\;\textrm{for } \lambda_i=\theta_{Z_i} + \beta^\top W_i.
\]
Prior models for $\theta_c$ and $\beta$ are as above. %For an example application see \cite{PHF12}.

\subsubsection{Extra variation in the response}
For some of the above models, it is possible that we may wish to allow for extra variation in the response. Our sampler is designed to achieve this by alternatively modelling $\lambda_i$ (as defined in the above response models) by
\[
\lambda_i=\theta_{Z_i}+\beta^\top W_i+\varepsilon_i,\;\;\;\textrm{where }\varepsilon_i\sim\mathrm{Normal}(0,\sigma^2_{\varepsilon}).
\]
Prior distributions for $\theta_c$ and $\beta$ are unchanged, but in this model $\Lambda$ contains an additional parameter, $\sigma^2_{\varepsilon}$, for which prior specification is required. For simplicity we work in terms of the precision $\tau_{\varepsilon}=1/\sigma^2_{\varepsilon}$, and adopt a gamma distribution with shape parameter $s_{\tau_{\varepsilon}}$ and rate parameter $r_{\tau_{\varepsilon}}$. This approach permits a simple Gibbs update of this parameter. In order to make inference about this model, it is also necessary to update the latent variables $\lambda_i$ at every sweep of the MCMC sampler. These parameters are considered an extension of $\Lambda$ (as they are not directly associated with a specific cluster) and are therefore updated in \emph{Step F} of the DPMM algorithm. Updates to these parameters are done using adaptive Random-walk-Metropolis steps.

\subsubsection{Gaussian response}
Our sampler is able to handle %Normally distributed 
continuous response data. As for many of the discrete response models, $\Theta_c$ is extended to contain $\theta_c$ for each $c$. As before $\Lambda$ contains $\beta$, but also $\sigma^2_Y$. These parameters allow us to write the response model as:
\[
f_Y(y_i|\bTheta_{Z_i},\Lambda,W_i)=p(Y_i|\theta_{Z_i},\beta,\sigma^2_Y,W_i)=\frac{1}{\sqrt{2\pi\sigma^2_Y}}\exp\left\{-\frac{1}{2\sigma^2_Y}(Y_i-\lambda_i)^2\right\},
\]
where $\lambda_i=\theta_{Z_i}+\beta^\top W_i$.

We impose the same prior settings as for the discrete response models, with the additional prior on $\tau_Y=1/\sigma^2_Y$ being $\mathrm{Gamma}(s_{\tau_Y},r_{\tau_Y})$, where $s_{\tau_Y}$ and $r_{\tau_Y}$ are the shape and rate hyper parameters that extend $\Theta_0$. Adopting this conjugate prior, updates for $\tau_Y$ are simple Gibbs updates.

\subsection{Variable selection}
\label{section:varselect1}
In addition to fitting mixtures, potentially linking covariates and responses, it may additionally be of interest to determine which covariates actively drive the mixture components, and which share characteristics common to all components. This can be formulated as a question of variable selection. Below we present details of how this idea can be modelled, first in the case of discrete covariates, as considered by \cite{PMH12}, and then in the context of Gaussian covariates, a formulation which, as far as we are aware, has not been reported elsewhere. Relevant to the variable selection formulation we have adopted is the model described in \cite{CD09}. A different modelling approach is presented in \cite{quintana13}. \cite{quintana13} consider the logit of the binary cluster specific selection switches and impose an additional level in the hierarchy using Normal densities and associated hyper-parameters. A normalization step is then required. In contrast, we impose an additional level in the hierarchy considering Bernoulli distributions for the binary switches, without the requirement of a normalization step. 

\subsubsection{Discrete covariates}
Following the approach taken by \cite{PMH12} our sampler implements two types of variable selection. We outline these approaches briefly in this section but for full details the reader is referred to this paper.

The first is a cluster specific variable selection approach, based on a modification of the model in \cite{CD09}.  Each mixture component $c$ has
an associated vector $\gamma_c = (\gamma_{c,1},\gamma_{c,2},\ldots,\gamma_{c,J})$, where $\gamma_{c,j}$ is a binary random variable that determines whether covariate $j$ is important to mixture component $c$. Let $\phi_{0,j,k}$  be the observed proportion of covariate $j$ taking the value $k$ throughout the whole covariate dataset $\bX$. Define the new composite parameters, 
\[
\phi^{*}_{c,j,k}:= \gamma_{c,j} \phi_{c,j,k} + (1-\gamma_{c,j}) \phi_{0,j,k} 
=\left(\phi_{c,j,k}\right)^{\gamma_{c,j}} \left(\phi_{0,j,k}\right)^{(1-\gamma_{c,j})}.  
\]
The above expression is substituted into Equation~\ref{eqn:discretemix} in place of $\phi_{c,j,k}$, to provide the likelihood for the covariate model. Under this model, each parameter vector $\Theta_c$ is extended by $\gamma_c$. We assume that, given $\rho_{j}$, the $\gamma_{c,j}$, $c=1,...,C$, are independent Bernoulli variables with $\gamma_{c,j} \sim \mbox{Bernoulli} (\rho_{j})$. We further consider a sparsity inducing prior for $\rho_{j}$ with an atom at zero, so that
\[
\rho_{j} \sim 1_{\{w_{j}=0\}} \delta_{0}(\rho_{j}) + 1_{\{w_{j}=1\}}  \mbox{Beta} (\alpha_{\rho}, \beta_{\rho}),
\]
where $w_{j} \sim \mbox{Bernoulli}(p_w)$. Therefore, additional parameters $\rho_j$ and $w_j$ are introduced into $\Lambda$. Here, $\alpha_{\rho}$ and $\beta_{\rho}$ are fixed, and can be specified by the user. The parameter $p_w$ is set equal to 0.5 by default, but it can also be specified by the user, also allowing for the atom at zero to be removed. The binary nature of $\gamma_{c,j}$ means that direct Gibbs updates can be used. This is an approach that allows for local cluster specific covariate selection, considering the $\gamma_{c,j}$ parameters, as well as global covariate selection, considering the overall selection probabilities $\rho_j$. 

An alternative approach to the variable selection presented above is a type of {\it soft} variable selection, where each covariate $j$, is associated with a latent variable $\zeta_j$, taking values in $[0,1]$, which informs whether variable $j$ is important in terms of supporting a mixture distribution. We define the new composite parameters as, 
\[
\phi^{*}_{c,j,k}:= \zeta_j \phi_{c,j,k} + (1-\zeta_j) \phi_{0,j,k}.
\]
which, as in the first variable selection model, is substituted into Equation~\ref{eqn:discretemix} in place of $\phi_{c,j,k}$, to provide the likelihood for the covariate model. Similarly to the first specification, we consider a sparsity inducing prior for $\zeta_{j}$ with an atom at zero, so that
\[
\zeta_{j} \sim 1_{\{v_{j}=0\}} \delta_{0}(\zeta_{j}) + 1_{\{v_{j}=1\}}  \mbox{Beta} (\alpha_{\zeta}, \beta_{\zeta}),
\]
where $v_{j} \sim \mbox{Bernoulli}(p_w)$. The parameter $p_w$ is set equal to 0.5 by default, but it can also be specified by the user, also allowing for the atom at zero to be removed. Conjugacy for $\phi_{c,j}$ is no longer retained, meaning that Metropolis-within-Gibbs updates are necessary. We use adaptive Random-walk-Metropolis proposals. The second alternative approach only allows for global variable selection and, in principle, is less likely to encounter mixing problems, compared to the more elaborate first formulation. Nevertheless, we have not yet observed considerable mixing problems when adopting either approach using PReMiuM.  For extended details of the conditional posteriors and updating strategy the interested reader is referred to \cite{PMH12}.

\subsubsection{Gaussian covariates}
\label{subsubsec:contvarselect}
The two variable selection methods described above can equally be applied to the Gaussian
mixture case. Defining $\bar{x}=(\bar{x}_1,\bar{x}_2,…,\bar{x}_J)$, where $\bar{x}_j=\frac{1}{n} \sum_i x_{i,j}$ is the average sample
value of covariate $j$, we can define the vector $\mu^{*}_{c}=(\mu^{*}_{c,1},\mu^{*}_{c,2},…,\mu^{*}_{c,J})$  where either
\[
\mu^{*}_{c,j} = \gamma_{c,j} \mu_{c,j} + (1-\gamma_{c,j}) \bar{x}_j
\]
or 
\[
\mu^{*}_{c,j} = \zeta_j \mu_{c,j} + (1-\zeta_j) \bar{x}_j 
\]
depending on which variable selection approach is being adopted. We then replace $\mu_{c,j}$  with $\mu^{*}_{c,j}$ in Equation~\ref{eqn:gaussmix}. Using identical priors for $\zeta_j$  or $\gamma_{c,j}$  and $\rho_j$ , these parameters are updated as for the discrete covariate case. The posterior conditional distributions for updating $\mu_c$  are
shown in Appendix \ref{appendix:contvarselect}, whereas those for updating other parameters are as before, with $\mu_{c,j}$ replaced with $\mu^{*}_{c,j} $ as appropriate.

\section{Predictions}\label{section:preds}

An important feature of our software is the computation of predicted responses for prediction scenarios. Suppose that we wish to understand the role of a particular covariate or group of
covariates. We can specify a number of predictive scenarios (or pseudo-profiles), that
capture the range of possibilities for the covariates that we are interested in. At each iteration the predictive subjects are assigned to one of the current clusters according to their covariate profiles. Seeing how these pseudo-profiles are allocated allows us to understand the risk associated with these profiles. 

The predictive subjects have no impact on the likelihood and so do not determine the clustering or parameters at each iteration and missing values in the predictive scenarios are ignored. At each sweep $r$ of the MCMC sampler we define an additional ``allocation'' variable, $\tilde{Z}^r_{s}$ corresponding to each predictive scenario $s$. Our software produces predicted values based on simple allocations or a Rao-Blackwellised estimate of predictions. The predicted values based on a simple allocation of cluster $c$ assign $\hat \theta^r_s = \theta^r_c$. For the Rao-Blackwellised predictions the probabilities of allocations are used instead of actually performing a random allocation. For each pseudo-profile we compute the posterior probabilities
$p(\tilde{Z}^r_s=c|\boldsymbol{x}_s,\boldsymbol{ \Theta } ^r , \boldsymbol{y},\boldsymbol{x}_1,\ldots,\boldsymbol{x}_N)$. With these probabilities we construct a cluster-averaged estimate of
$\theta$ for each particular pseudo-profile at each sweep. Specifically,
\[
 \hat{\theta}^r_{p}=\sum_{c=1}^{\infty}
p(\tilde{Z}^r_{p}
=c|\boldsymbol{x}_p,\boldsymbol{\Theta}^r,\boldsymbol{y},\boldsymbol{x}_1,\ldots,\boldsymbol{x}_N)\theta^r_c.
\]
Looking at the density of these predictions over MCMC sweeps gives us an estimate
of the effect of a particular pseudo-profile, and its comparison to other
pseudo profiles, allowing us to derive a better understanding of the role of
specific covariates. Moreover, the impact of ignoring missing values in the pseudo-profiles essentially means that the missing value will reflect the covariate patterns 
present in the main sample. Because of this, the marginal effect of covariates or groups of covariates that is derived has to be interpreted as a population average effect, over a population with similar characteristics to that under study.

If a subject is missing fixed effects, then the mean value or 0 cateory fixed effect is used in the predictions. In this case, effectively, the fixed effects do not contribute to the predicted response.  If the offset or number of trials is missing, this value is taken to be 1 when making predictions.
 
\section{Postprocessing of the MCMC output}\label{sec:postprocessing}

The rich posterior output produced can be used to learn about the partition space and its uncertainty. It is useful to show a representative partition, as an effective way to convey the output of the clustering
algorithm. Moreover, it is also of interest to assess the uncertainty associated with subgroups of this best partition. 

We discuss below the necessary steps. See also \cite{MPJ10}.

\begin{enumerate}
\item \textit{Computation of the dissimilarity matrix}.  Due to the problem of ``label switching'', i.e the labels associated with each cluster change during the MCMC iterations, we can not simply assign each observation to the cluster that maximises the average posterior probability. Methods that deal with label switching, like the relabeling algorithm of \cite{Stephens00}, require the number of clusters $K$ to be fixed. Using the Dirichlet process mixture models, we allow the number of clusters to vary from one MCMC sample to the next. One possible solution is to choose the partition based on a posterior similarity matrix. At each iteration of the sample, we record pairwise cluster membership and construct a score matrix, with entries equal to 1 for pairs belonging to the same cluster and 0 otherwise. Averaging these matrices over the whole MCMC run leads to a similarity matrix $S$, which can be then used to identify an optimal partition.

%One possible solution is to choose a partition based on a posterior similarity matrix $P(c_{i}=c_{j}|y)$, which is an $n \times n$ matrix that contains the pairwise probabilities that two observations belong to the same cluster.  This similarity matrix $S$ can be then used to identify an optimal partition.
%Different approaches based on the posterior similarity matrix or not have been proposed in order to derive the representative clustering.

\item \textit{Identifying the optimal partition.} Many methods to identify the optimal partition using the posterior similarity matrix have been proposed in the literature. The similarity matrix computed by \pkg{PReMiuM} can be processed using the \proglang{R} package \pkg{mcclust} \citep{Fritsch09}. We have implemented directly in \pkg{PReMiuM} two deterministic clustering procedures to characterise the optimal partition. 

The first finds the best partition by choosing the one which minimises the least-square distance to the matrix $S$. This approach is equivalent to the Binder's loss method \citep{Fritsch09}. It is fast, but in our experience it is susceptible to Monte Carlo error. 

The second procedure implemented in the package is Partitioning Around Medoids (PAM) on the dissimilarity matrix $1-S$. PAM is available in \proglang{R} in the package \pkg{cluster} and it robustly assigns individuals to clusters in a way consistent with matrix $S$. PAM is implemented for each possible number of clusters up to a specified maximum, and for each fixed number of clusters the best PAM partition is selected. A final representative cluster is then chosen by maximising the average silhouette width across these best PAM partitions.

\item \textit{Computation of the average risk and profile and the corresponding credible intervals.} Given an optimal partition $P^\ast$ obtained as above, we examine the MCMC output to assess whether or not the model consistently clusters individuals in a manner similar to $P^\ast$. 

For example, for Bernoulli response, we obtain a distribution of the baseline risks for each cluster defined by $P^\ast$. At each iteration of the sampler we compute the average of baseline risks $p_{z_i}$, defined in Section \ref{section:profile}, for all individuals within a particular cluster $k$ of the optimal partition. This average baseline risk for cluster $k$ is computed as follows: 
\begin{equation*}
\bar p _k = \frac{1}{n_k} \sum_{i: z_i^{P^\ast}=k} p_{z_i}
\end{equation*}
where $n_k$ denotes the number of individuals in cluster $k$. This provides an empirical sample from the baseline risk associated with cluster k. Consistent clustering leads to narrower credible intervals derived from this distribution. In a similar way we can compute the distribution of cluster parameters for other response and covariates types.
\end{enumerate}

\section{Mixing of the MCMC algorithm}
\label{sec:mixing}

The likelihood of the DPMM is invariant to the order of cluster labels but the prior specification of the stick breaking construction is not. Therefore, to ensure adequate mixing across orderings, it is important to include label-switching moves. In this package we have implemented the two label switching moves proposed by \cite{PR08} as well as a third label switching move proposed by \cite{HLR13}. This latter move updates the cluster weights so that for each cluster being updated, the proposed new weight is the expected value of the weight conditional upon the new allocations, adjusted by the ratio of the existing weight and its expected value conditional upon the existing allocations, with the weights appropriately renormalised. See Equation~6 of \cite{HLR13} for details of the move, and more generally for a review of the sampler performance.

Even with these label switching moves, convergence may be problematic and the user must address this issue using diagnostic tools. One difficulty in this respect is that there are no parameters in the model that can reliably demonstrate convergence. The parameters of the fixed effects tend to converge very quickly, regardless of the underlying clustering, as they are not cluster specific and therefore are not a good indication of the overall convergence. Plotting functions to assess convergence of the global parameters are included in the package and are discussed in Section \ref{sec:example}. The cluster parameters, such as the $\theta_c$'s, cannot be tracked as their number (and their labels) can change from one iteration to the next. The concentration parameter $\alpha$ is not a reliable indicator of convergence either, as discussed in \cite{HLR13}. 

To overcome this challenge, we have implemented the computation of the marginal model posterior $p(\mathbf{Z}|\mathbf{D})$ as an additional diagnostic tool. This represents the posterior distribution of the allocations given the data, having marginalised out all the other parameters \citep{HLR13}. The marginal model posterior is computed for each run of the MCMC and it has proved very effective for our real examples to compare runs with different initialisations and identify runs that were significantly different from others. Our experience suggests that it is harder for the MCMC algorithm to split rather than merge clusters. This means that it is important to initialise the algorithm with a number of cluster which is greater than the number of clusters that the algorithm will convergence to. The marginal model posterior can help to assess what such number is for each specific example.

Finally, while optimal partitions allow visualisation of the result of a clustering algorithm, such an approach must be applied with care as we are not aware of any effective method to directly quantify nor visualise clustering uncertainty. For this reason, we advise using predictions as an additional tool to assess convergence and visualise the output of the algorithm, as their posterior distributions can be compared across runs using standard methods. More details of using the package for predictions can be found in Section \ref{sec:example}. We have observed that these posterior predictive distributions tend to be more stable than optimal partitions.

\section{Software}
\label{sec:software}
Our implementation of the DPMM algorithm is available as an \proglang{R} package from CRAN. The software is primarily written in  \proglang{C++} and  \proglang{R}.
%, and is based on the MCMCpp template  \proglang{C++} header library for MCMC samplers \todo{Add reference}. The model also 
%and it makes use of the Boost \proglang{C++} header library \citep{Boost}, along with the Eigen \proglang{C++} header library \citep*{Eigen} for linear algebra (needed specifically for models with Gaussian covariates).

The sampler implements the algorithm exactly as detailed in the current paper, although continued work is in progress to extend the scope of the software to cover additional models. 
%At run time the user provides a set of command line arguments to specify which model they wish to run, the data files to be used, the number of sweep to run. A full list of options (with their default values is can be used by running the program with the ``--help'' argument. 

The program is further customisable through the specification of hyperparameters, providing name-value pairs for the various hyperparameters used within the model being run. If the value is not set for a specific hyperparameter, it takes its default value. Default values can be found within the full documentation that is available as part of the software and an example in provided in Section~\ref{code:varselect}. 
 
Moreover, this package can produce predicted values based on random allocations, or a Rao-Blackwellised estimate of predictions, where the probabilities of allocations are used instead of actually performing a random allocation.

%The software also includes an accompanying \proglang{R} package, designed primarily for limited post processing and graphical visualisation of the \proglang{C++} output, of the sort described by \cite{MPJ10} and \cite{PMH12}.

\section{Examples}
\label{sec:example}

\subsection{Simulated example}\label{sect:simulation}

We simulated 1,000 subjects, partitioned into 5 groups in a balanced manner. Ten binary covariates were considered. To demonstrate one of the variable selection approaches within the sampler, only the first eight covariates support a clustering structure. The response is binary. This dataset can be simulated as follows.

\begin{CodeChunk}
\begin{CodeInput}
R> library("PReMiuM")
R> inputs <- generateSampleDataFile(clusSummaryVarSelectBernoulliDiscrete())
\end{CodeInput}
\end{CodeChunk}

We use the default values for all hyperparameters:  Dirichlet conjugate priors with $a_j=1$ for the covariates and
\begin{eqnarray*}
p(\alpha)&\equiv& \mbox{Gamma}(2,1)\\
p(\theta_c) &\equiv& \mbox{t}_7(0,2.5)\\
p(\beta)&\equiv&\mbox{t}_7(0,2.5)
\end{eqnarray*}
where $\mbox{Gamma}(\alpha,\beta) \equiv \frac{\beta^\alpha}{\Gamma(\alpha)} x^{\alpha \,-\, 1} e^{- \beta x }$ and  
\begin{equation}
\mbox{t}_\nu (\mu,\sigma) \equiv  \frac{\Gamma(\frac{\nu + 1}{2})}{\Gamma(\frac{\nu}{2})\sqrt{\pi\nu}\sigma} \left[1+\frac{1}{\nu}\left(\frac{x-\mu}{\sigma}\right)^2\right]^{-\frac{\nu+1}{2}}.
\end{equation}

We initialised all chains allocating subjects randomly to 20 groups. We run the chain for 10,000 after a burn-in sample of 20,000 iterations. While ensuring convergence is a complex problem, we have observed good stability in all our runs, with results from independent chains virtually identical. 

\begin{CodeChunk}
\begin{CodeInput}
R> runInfoObj <- profRegr(yModel=inputs$yModel, xModel=inputs$xModel, 
    nSweeps=10000, nBurn=20000, data=inputs$inputData, 
    output="output", covNames=inputs$covNames, nClusInit=20, run=TRUE)
R> dissimObj <- calcDissimilarityMatrix(runInfoObj)
R> clusObj <- calcOptimalClustering(dissimObj)
R> riskProfileObj <- calcAvgRiskAndProfile(clusObj)
R> clusterOrderObj <- plotRiskProfile(riskProfileObj, 'summary-sim.png') 
\end{CodeInput}
\end{CodeChunk}

\begin{figure}
\centering
\includegraphics[width=21cm,angle=270]{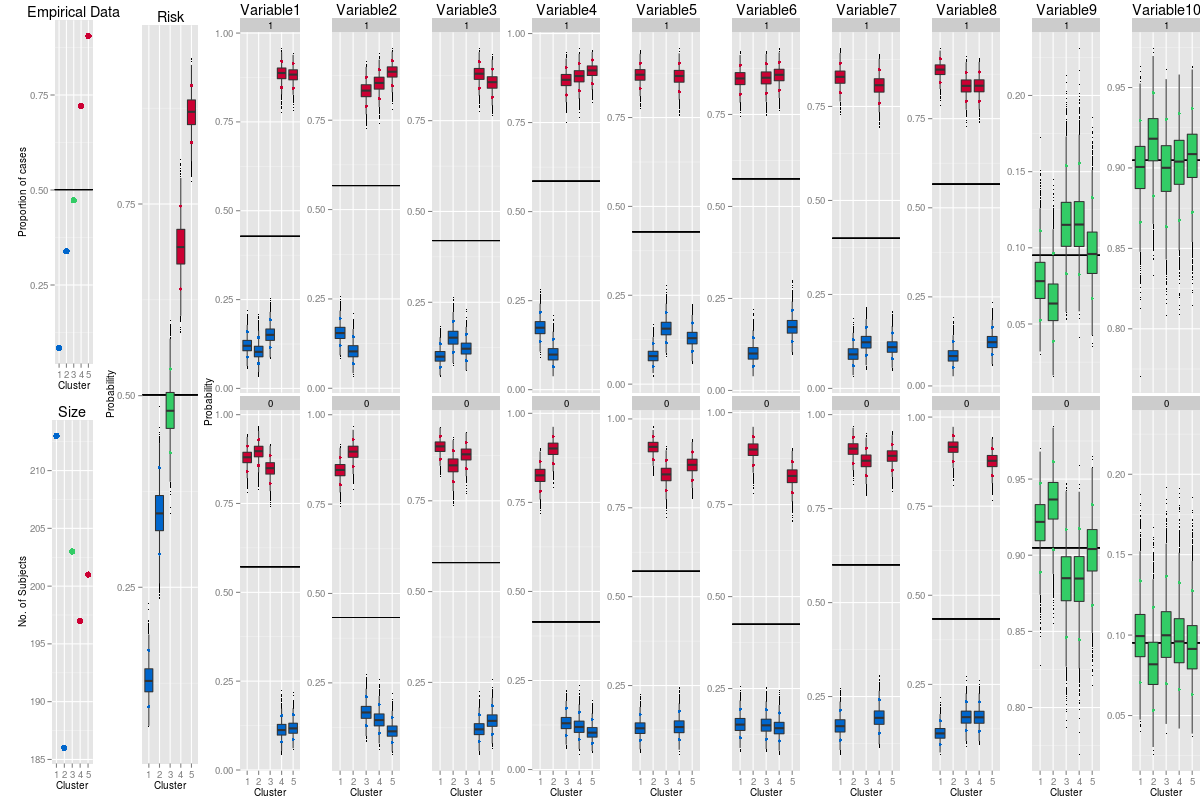}
\caption{Posterior distributions of the parameters for binary response and discrete covariates for the representative clustering.\label{fig:sim-risk}}
\end{figure}

Figure~\ref{fig:sim-risk} shows a box-plot of the posterior distribution for the probabilities of the response and the covariates for the 5 clusters that form the representative clustering. Additionally, the package includes the function \code{heatDissMat()} which produces a heatmap of the dissimilarity matrix, rearranged such that observations with high pairwise cluster membership appear consecutively.  

\subsection{Predictions}
\begin{figure}
\centering
\includegraphics[width=7cm]{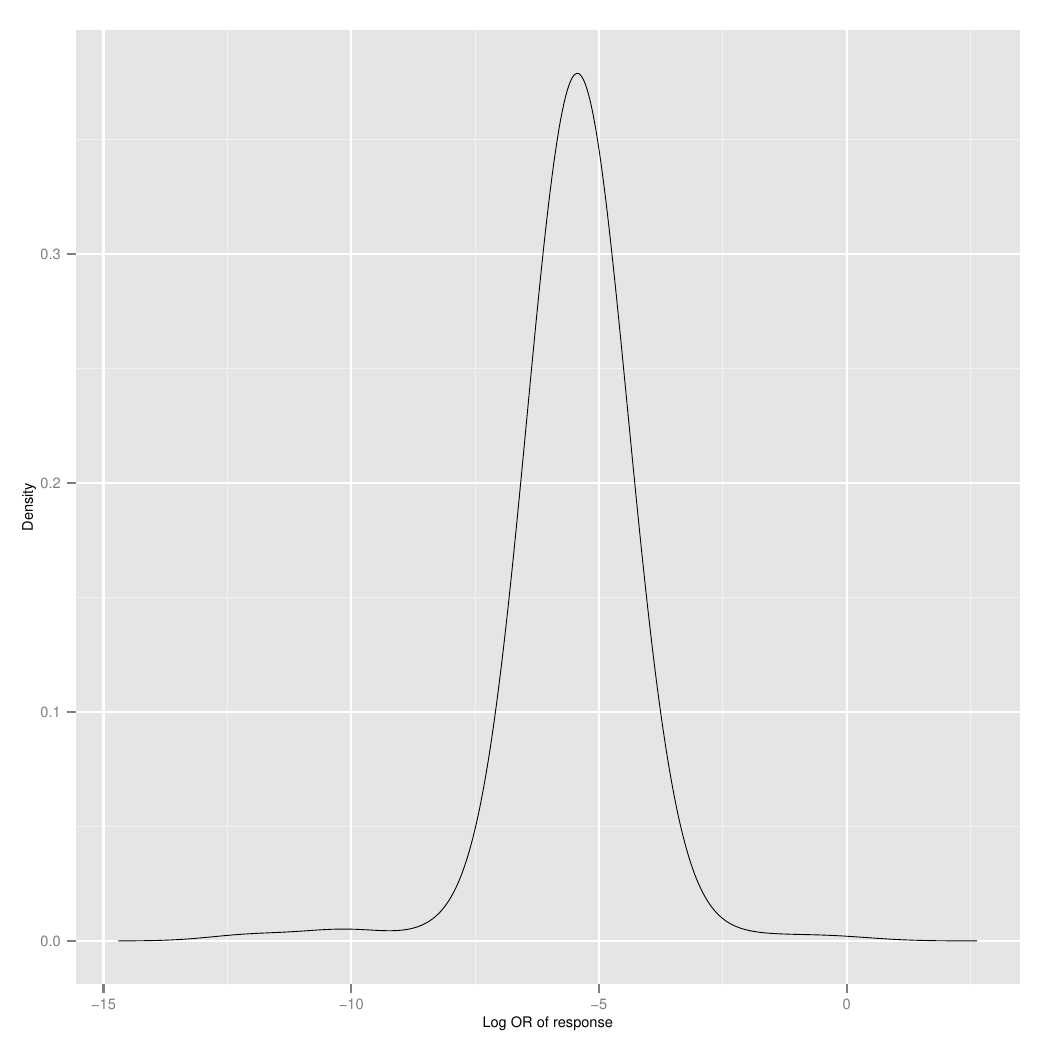}
\includegraphics[width=7cm]{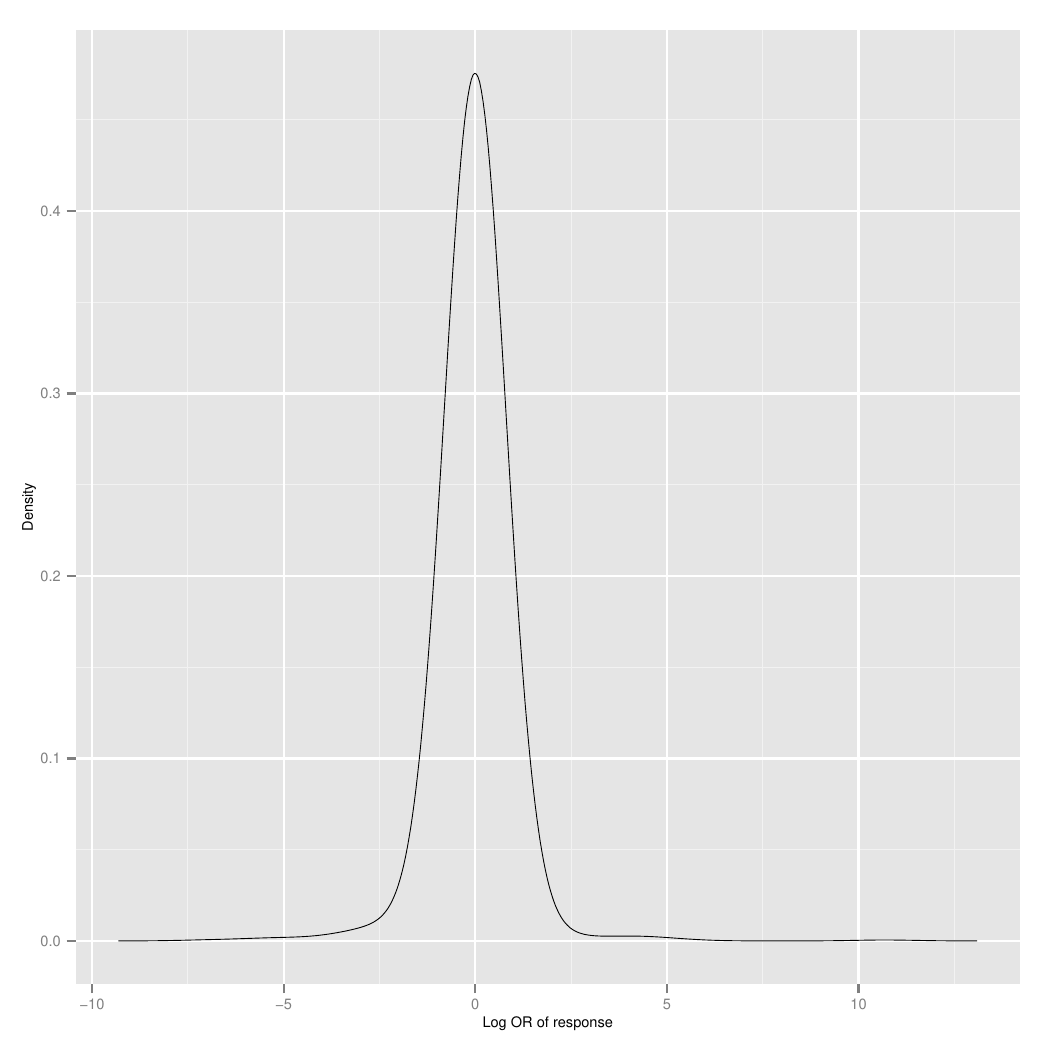}
\caption{The predictive distribution of the response for two prediction scenarios. The covariate values for the predictive scenarios are [2,2,2,2] and [0,0,NA,0,0] respectively. `NA' represents a missing value.\label{fig:predictions}}
\end{figure}

\pkg{PReMiuM} can produce predicted values based on simple allocations (the default), or a Rao-Blackwellised estimate of predictions, where the probabilities of allocations are used instead of actually performing a random allocation. The following code can be used to reproduce the predictive distribution plotted in Figure~\ref{fig:predictions}. As discussed in Section~\ref{section:preds}, the missing values, as in the second prediction scenario given below, are ignored and their marginal effect can be interpreted as a population average effect. The predictions are consistent with the simulated data. 

\begin{CodeChunk}
\begin{CodeInput}
R> inputs <- generateSampleDataFile(clusSummaryBernoulliDiscrete())
R> preds<-data.frame(matrix(c(
    2, 2, 2, 2, 2,
    0, 0, NA, 0, 0),ncol=5,byrow=TRUE))
R> colnames(preds)<-names(inputs$inputData)[2:(inputs$nCovariates+1)]
R> runInfoObj<-profRegr(yModel=inputs$yModel, xModel=inputs$xModel, 
    nSweeps=1000, nBurn=1000, data=inputs$inputData, output="output", 
    covNames=inputs$covNames,predict=preds,
    fixedEffectsNames = inputs$fixedEffectNames)        
R> dissimObj <- calcDissimilarityMatrix(runInfoObj)
R> clusObj <- calcOptimalClustering(dissimObj)
R> riskProfileObj <- calcAvgRiskAndProfile(clusObj)
R> predictions <- calcPredictions(riskProfileObj,fullSweepPredictions=TRUE,
    fullSweepLogOR=TRUE)
R> plotPredictions(outfile="predictiveDensity.pdf", runInfoObj=runInfoObj,
    predictions=predictions, logOR=TRUE)
\end{CodeInput}
\end{CodeChunk}

\subsection{Variable selection}\label{code:varselect}

Note that covariates 9 and 10 in Figure~\ref{fig:sim-risk} have similar profile probabilities for all clusters, as they have been simulated not to affect the clustering. The variable selection approach will identify the covariates that do not contain clustering support and exclude them from affecting the clustering. 

\begin{figure}
\centering
\includegraphics[width=14cm]{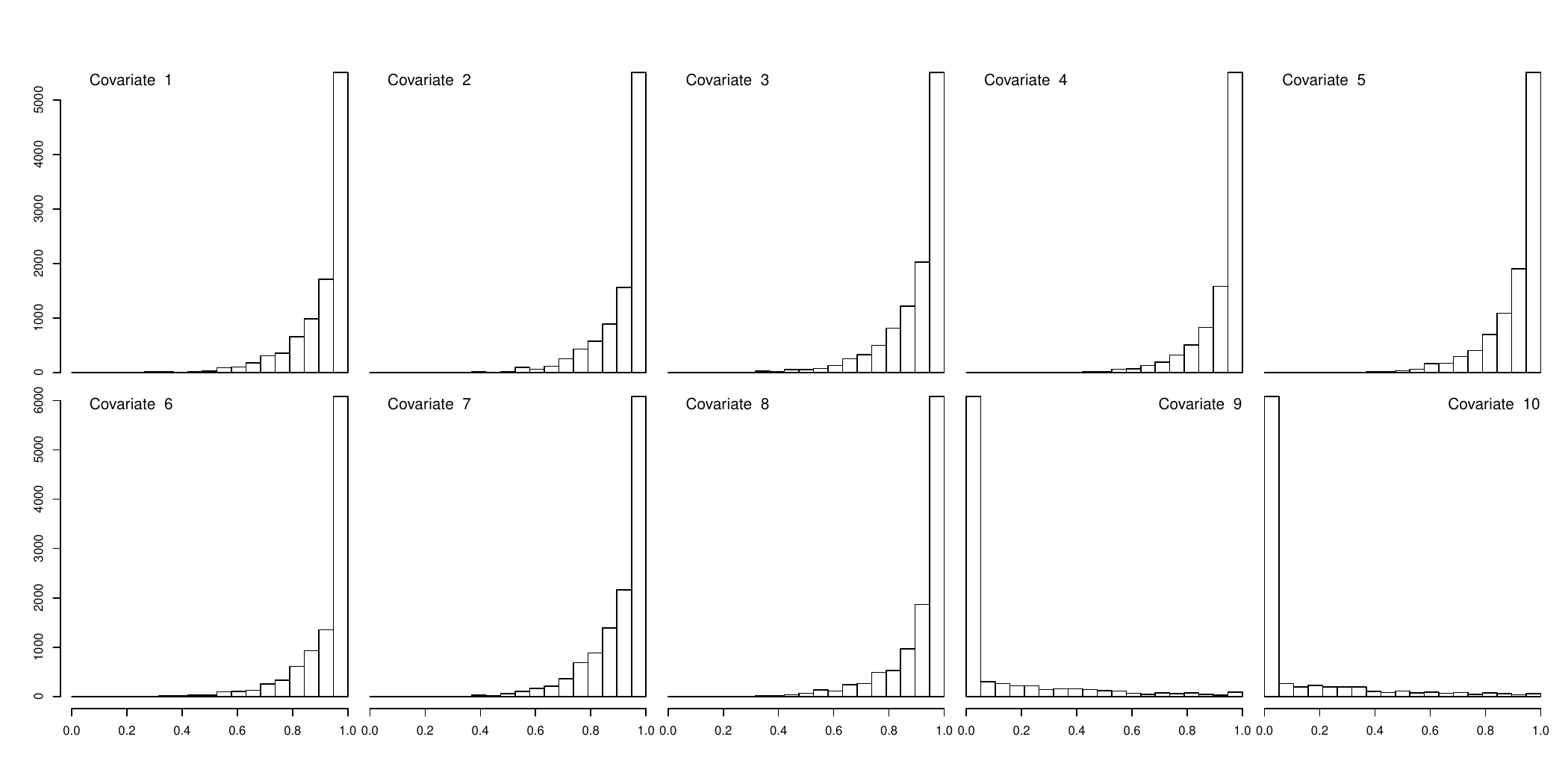}
\caption{The sampled values of the binary variable $\rho_j$, used for variable selection, for each covariate.\label{fig:varSelect}}
\end{figure}

We initialised the chains as in the simulated example above, with additional prior specifications given by
\[
\rho_{j} \sim 1_{\{w_{j}=0\}} \delta_{0}(\rho_{j}) + 1_{\{w_{j}=1\}}  \mbox{Beta} (0.5, 0.5),
\]
where $w_{j} \sim \mbox{Bernoulli}(0.5)$.
The algorithm consistently sampled values $\rho_p$ in accordance with the simulated data, as shown in Figure~\ref{fig:varSelect}. This figure can be reproduced as follows.

\begin{CodeChunk}
\begin{CodeInput}
R> inputs <- generateSampleDataFile(clusSummaryVarSelectBernoulliDiscrete())
R> hyp<- setHyperparams(aRho=0.5, bRho=0.5, atomRho=0.5)
R> runInfoObj<-profRegr(yModel=inputs$yModel, xModel=inputs$xModel, 
    nSweeps=10000, nBurn=10000, data=inputs$inputData, 
    output="output", covNames=inputs$covNames, 
    varSelectType="BinaryCluster", hyper=hyp)

R> rho <- summariseVarSelectRho(runInfoObj)
R> par(mfrow=c(5,2))
R> for (k in 1:runInfoObj$nCovariates){	
R>     hist(rho$rho[,k], xlim=c(0,1), main="")
R> }
\end{CodeInput}
\end{CodeChunk}

\subsection{Assessing convergence}

There is no method that can assure us that our MCMC chains have converged to the posterior probability distribution but there are several methods that can investigate whether there is evidence against convergence. 

We have implemented the function \code{globalParsTrace()} which plots the trace of some global parameters such as $\alpha$, $\beta$ and the number of clusters. For more convergence diagnostics the samples for all global parameters can be analysed using the \proglang{R} package \pkg{coda}.
\pkg{coda} is an \proglang{R} package to perform convergence diagnostics and statistical and graphical output analysis of the output from an MCMC sampler. 

\begin{figure}
\centering
\includegraphics[width=7cm]{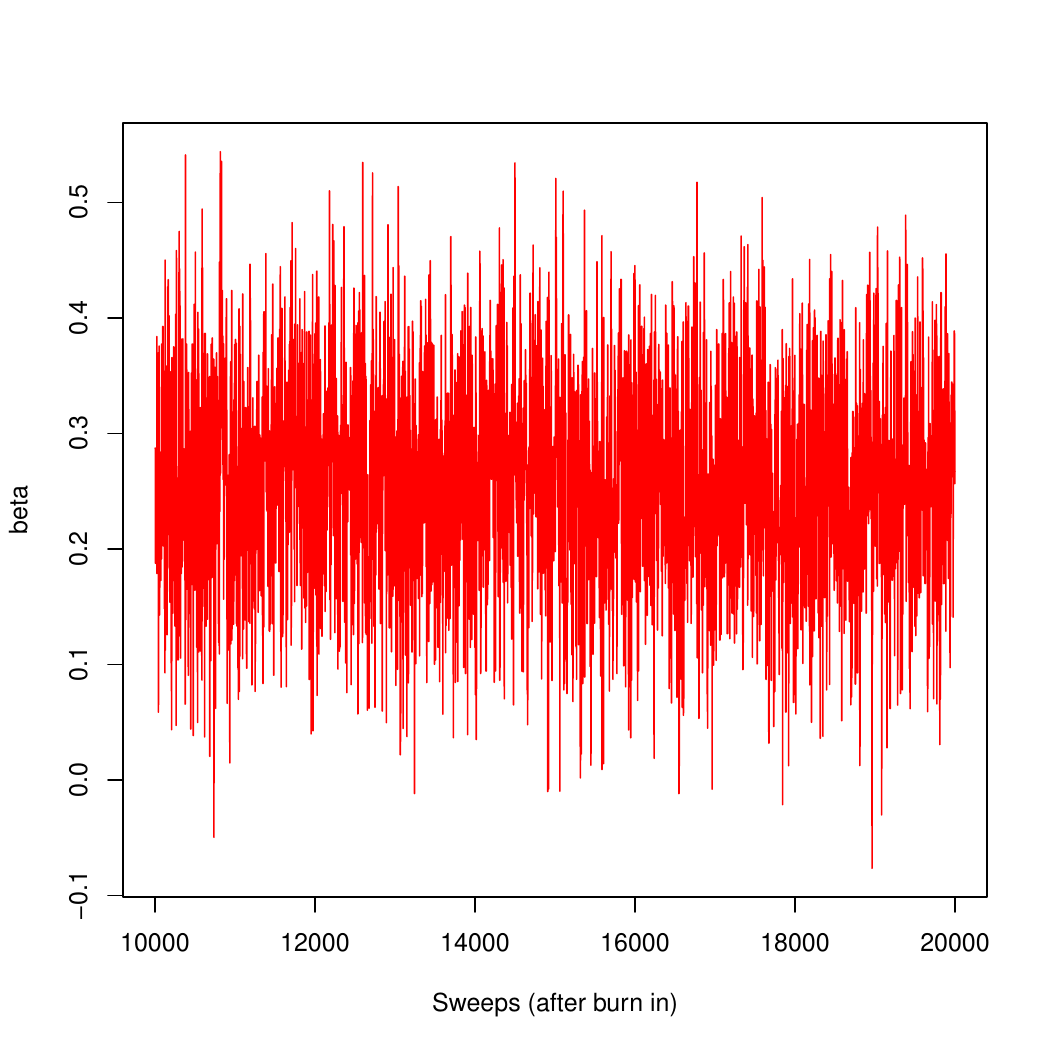}
\includegraphics[width=7cm]{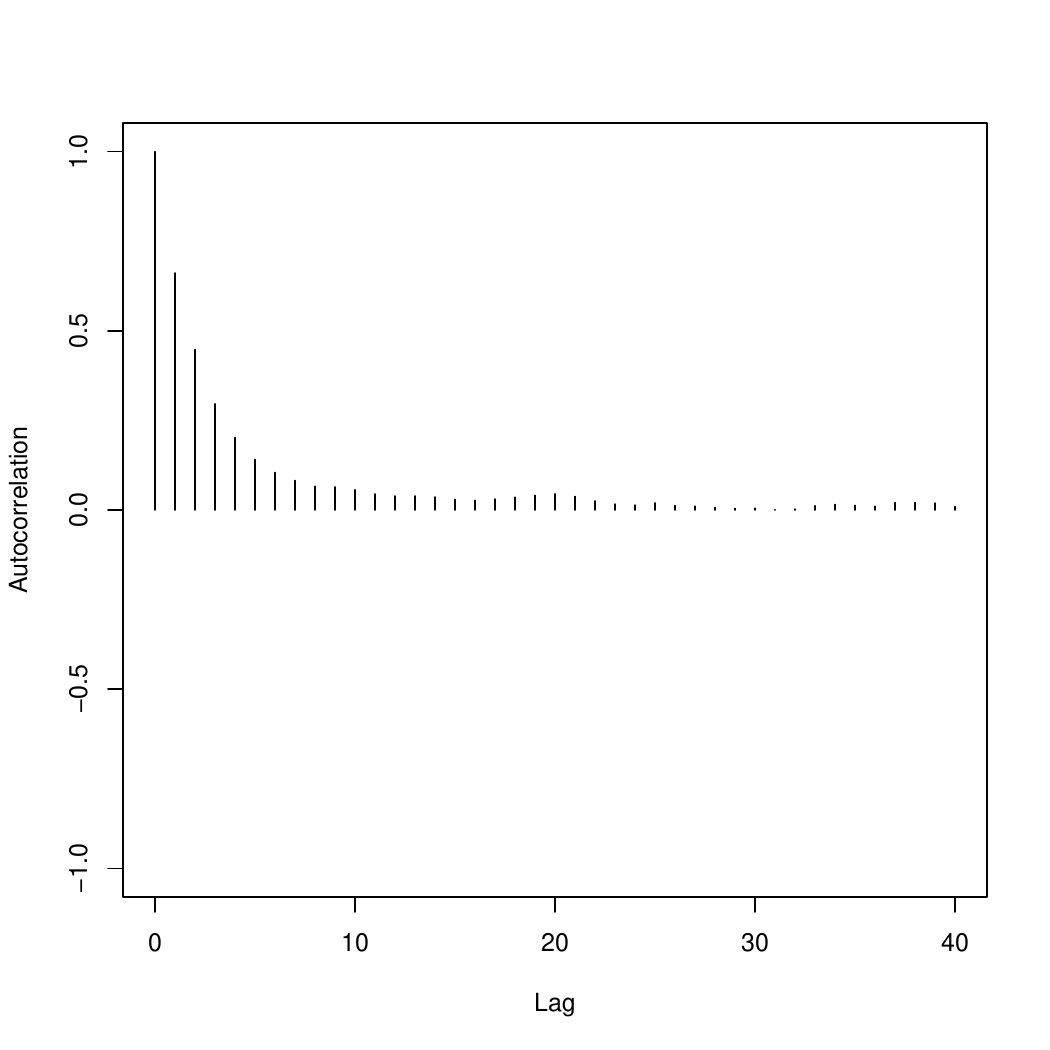}
\caption{Convergence diagnostics for parameter $\beta_1$: trace plot and autocorrelation plot done using \pkg{coda}. \label{fig:conv}}
\end{figure}

The following code can be used to reproduce the trace plot and autocorrelation plot in Figure~\ref{fig:conv} for parameter $\beta_1$. 

\begin{CodeChunk}
\begin{CodeInput}
R> inputs <- generateSampleDataFile(clusSummaryVarSelectBernoulliDiscrete())
inputs <- generateSampleDataFile(clusSummaryBernoulliDiscrete())
R> runInfoObj<-profRegr(yModel=inputs$yModel, xModel=inputs$xModel, 
    nSweeps=10000, nBurn=10000, data=inputs$inputData, 
    output="output", covNames=inputs$covNames,
    fixedEffectsNames = inputs$fixedEffectNames)

R> globalParsTrace(runInfoObj,parameters="beta",plotBurnIn=FALSE, whichBeta=1)
R> library(coda)
R> betaChain<-mcmc(read.table("output_beta.txt")[,1])
R> autocorr.plot(betaChain)
\end{CodeInput}
\end{CodeChunk}

The cluster specific parameters cannot be plotted as easily due to label switching and assessing their convergence is not an easy task. \cite{HLR13} introduce the marginal model posterior as a tool to assess convergence for Dirichlet process mixtures. We define the marginal partition posterior as $p(\mathbf{Z}|\mathbf{D})$. This quantity represents the posterior distribution of the allocations given the data, having marginalised out all the other parameters.

\begin{figure}
\centering
\includegraphics[width=8cm]{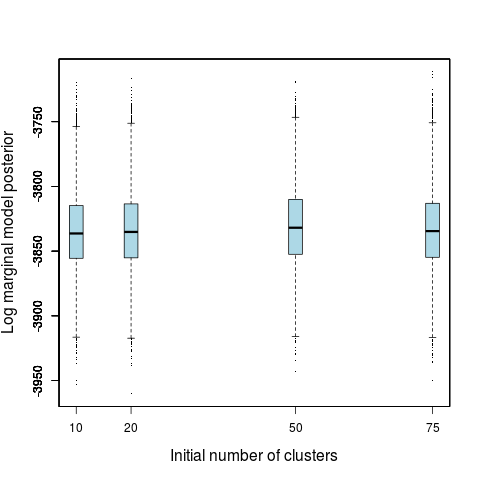}
\caption{The log marginal model posterior for four runs of profile regression, on the same dataset but with different initialisations (ie., different initial number of clusters). \label{fig:marginalModelPosterior}}
\end{figure}

The marginal model posterior can be computed in \pkg{PReMiuM}. The code below computes the marginal model posterior for four different runs of profile regression on the same dataset with different initialisations - different number of initial clusters. As seen in Figure~\ref{fig:marginalModelPosterior}, plotted using the code below, for the given simulated dataset, all the MCMC runs appear to converge to subsets of the model space with equivalent marginal model posterior. This does not imply convergence, but it is a useful diagnostic tool as it can highlight a lack of convergence in certain circumstances \citep{HLR13}. The marginal model posterior can also be plotted using the function \code{globalParsTrace()}.

\begin{CodeChunk}
\begin{CodeInput}
R> inputs <- generateSampleDataFile(clusSummaryBernoulliDiscrete())
R> nClusInit<-c(10,20,50,75)
R> for (i in 1:length(nClusInit)){
R> runInfoObj<-profRegr(yModel=inputs$yModel, 
    xModel=inputs$xModel, nSweeps=10000,
    nBurn=10000, data=inputs$inputData, 
    output=paste("init",nClusInit[i],sep=""), 
    covNames = inputs$covNames, alpha=1, 
    fixedEffectsNames = inputs$fixedEffectNames,nClusInit=nClusInit[i])
R>   margModelPosterior(runInfoObj)
R> }
R> mmp<-list()
R> for (i in 1:length(nClusInit)){
R> 	mmp[[i]]<-read.table(paste("init",nClusInit[i],
             "_margModPost.txt",sep=""))[,1]
R> }
R> plot(c(head(nClusInit,n=1)-0.5,tail(nClusInit,n=1)+0.5),
    c(min(unlist(mmp)),max(unlist(mmp))),type="n",
    ylab="Log marginal model posterior",
    xlab="Initial number of clusters",
    cex.lab=1.3,xaxt="n")
R> axis(1,at=nClusInit,labels=nClusInit)
R> for (i in 1:length(nClusInit)){
R> boxplot(mmp[[i]],add=T,at=nClusInit[i],pch=".",
    boxwex=5,col="lightblue")
R> }
\end{CodeInput}
\end{CodeChunk}

\subsection{Run times}
We have run simulations to test  \pkg{PReMiuM}'s speed and how it scales when the number of subjects or covariates increases. The code was run in serial on an Intel(R) Xeon(R) CPU E5-2650 clocked at 2.00GHz with 20MB L3 cache, on a system with 64GB RAM. 

\begin{table}[h]
\centering
\caption{Time to run 100 iterations of  \pkg{PReMiuM} for Bernoulli response and Discrete covariates}
\begin{tabular}{c|ccc}
&\multicolumn{3}{c}{\bf{Number of covariates}}\\
\hline
\bf{Number of Subjects} &100&1,000&10,000\\
\hline
1,000&   4 sec&1 min& 13 min\\
2,500&11 sec& 1.4 min&26 min\\
5,000&32 sec&3.5 min &32 min\\
\end{tabular}
\end{table}

\begin{table}[h]
\centering
\caption{Time to run 500 iterations of  \pkg{PReMiuM} for continuous response and continuous covariates}
\begin{tabular}{c|cc}
&\multicolumn{2}{c}{\bf{Number of covariates}}\\
\hline
\bf{Number of Subjects} &50&100\\
\hline
250&34 sec&4 min\\
500&51 sec&5 min\\
1,000&1.3 min&7 min\\
\end{tabular}
\end{table}

\section*{Conclusions}
\label{sec:conclusions}
The structure of \pkg{PReMiuM} objects gives rise to a wider variety of uses than can be described in detail here. Our intention was to provide a tutorial for Dirichlet process clustering and to illustrate the basic features of the sampler and post-processing tools that we have implemented in \pkg{PReMiuM} to demonstrate its utility. Our long-term goal is to continue to develop this package for analysis on complex and high dimensional datasets as well to increase the flexibility with regards to the data types that can be analysed. 

\section*{Acknowledgements}
Silvia Liverani acknowledges support from the Leverhulme Trust (ECF-2011-576). David I. Hastie acknowledges support from the INSERM grant (P27664). We are grateful for
helpful discussions with Sara K. Wade.

\begin{appendix}
\section{Appendices}
\subsection{}
\label{appendix:cstar}
We provide the following proposition to support our assertions regarding $C^{\star}$.

\begin{proposition}
Suppose that we have a model with posterior as given in Equation~\ref{eqn:dpmmpostu}. Suppose $Z^{\star}$,$U^{\star}$ and $C^{\star}$ are defined as in Section \ref{subsec:blocksampler}. Then:
\begin{enumerate}
   \item[(i)] $\psi_c<U_i\;$ for all $i=1,2,\ldots,n$ and all $c>C^{\star}$ almost surely;
   \item[(ii)] $C^{\star}\geq Z^{\star}$ almost surely; and
   \item[(iii)] $C^{\star}<\infty$ almost surely.
\end{enumerate}
\begin{proof}
We rely on the fact that if $V_1,V_c\sim\mathrm{Beta}(1,\alpha)\;$,  $\psi_1=V_1\;$ and $\psi_c=V_c\prod_{l<c}(1-V_c)\;$ for $c=2,3,\ldots$ then $\sum_{c=1}^{\infty}\psi_c =1$. A proof of this result for the DPMM (in terms of more general conditions) is provided by \cite{IJ01}. Then:
\begin{itemize}
\item[\emph{(i)}] By definition, for all $i=1,2,\ldots,n$ 
\[
U_i \geq U^{\star} > 1 - \sum_{c=1}^{C^{\star}}\psi_c =  \sum_{c=C^{\star}+1}^{\infty}\psi_c  \geq \psi_{c'} \;\;\forall {c'}>C^{\star}. 
\]
\item[\emph{(ii)}] Let $i^{\star}$ be an individual $i$ such that $Z_i=Z^{\star}$. Again, by definition, $U^{\star}\leq U_{i^{\star}} < \psi_{Z^{\star}}$. This implies $1-\psi_{Z^{\star}} < 1-U^{\star}$, meaning
\[
   \sum_{c=1}^{Z^{\star}-1}\psi_c < 1-\psi_{Z^{\star}} < 1 - U^{\star}.
\]
By definition of $C^{\star}$, this implies $C^{\star} \geq Z^{\star}$ almost surely.
\item[\emph{(iii)}] Since $\sum_{c=1}^{\infty}\psi_c =1$, this is a convergent series. By definition of a convergent series and because $U^{\star}>0$ we have $C^{\star}<\infty$ almost surely.
\end{itemize}
\end{proof}
   
\end{proposition}

\subsection{}
\label{appendix:algo1}
Below are additional comments to explain the blocking strategy employed in the DPMM algorithm. We use `$\cdot$' to denote ``all other parameters and data''.
\begin{description}
 \item[\emph{Step A.}] This step is a straightforward calculation of $Z^{\star}$, which (potentially) changes at each iteration (with the update of $\bZ$). The set $A$ is defined immediately conditional on this value.
 \item[\emph{Step B.}] This is a joint update of $\bU$ and the parameters corresponding to the active components in $A$, with the inclusion of label switching moves. The principle is to use the identity $p(\bV^A,\bTheta^A,\bZ,\bU|\cdot)=p(\bV^A,\bTheta^A,\bZ|\cdot)p(\bU|\bV^A, \bTheta^A,\bZ,\cdot)$. We proceed by first updating $(\bV^A,\bTheta^A)\sim p(\bV^A,\bTheta^A|\bZ,\cdot)=\int p(\bV^A,\bTheta^A,\bU|\bZ,\cdot)\mathrm{d}\bU.$ Due to the conditional independence of $\bV$ and $\bTheta$, this can be done in two steps: updating $\bV$ (\emph{B.1}) then updating $\bTheta$ (\emph{B.2}). The moves are presented as Gibbs updates, but in fact they are Metropolis-Hastings moves, where the variable of interest (for example $\bV^A$) is sampled from its full conditional, and the other variables (for example $\bTheta^A$ and $\bZ$) are kept fixed. This results in an acceptance probability of 1, making the Gibbs update equivalent. The updated values are then used as interim values for $\bV$ and $\bTheta$ in (Metropolis-Hastings) label-switching moves (see Section \ref{sec:mixing}) which are applied in \emph{B.3}. The moves can change the values of $\bV$ as well as their order. The resulting sample of $\bV$ and $\bTheta$ are the final updated values of these parameters for this sweep. The updated allocation vector $\bZ$ is used as an interim value throughout the remainder of the steps of the sweep, before the final updated value of $\bZ$ is sampled in \emph{Step G}. The final part (\emph{B.4}) of \emph{Step B} is to update $\bU$ conditional upon the updated value of $\bV$ and $\bTheta$ and the interim value of $\bZ$.
 \begin{description}
  \item[\emph{B.1}] Integrating out $\bU$ and taking advantage of the conjugacy of the distribution for $\bV$ inherent in the DPMM, along with the conditional independence structure, each component of vector $\bV^A$ is updated by sampling $V_c\sim\mathrm{Beta}(1-d+n_c,\alpha+dc+n^+_c)$, $c\in A$, where $n_c=\sum_i\boldsymbol{1}_{\{Z_i=c\}}$ and $n^+_c=\sum_i\boldsymbol{1}_{\{Z_i>c\}}$. The Dirichlet process is a special case of the Pitman-Yor process for $d=0$. 
  \item[\emph{B.2}] Integrating out $\bU$ and taking advantage of the conditional independence structure, $\bTheta^A$ is updated from $p(\bTheta^A|\bZ,\Theta_0,\bD)$. The full details of this will depend upon the application and the choice of $f$ and $P_{\Theta_0}$. Examples are given in Sections \ref{sec:models}. \item[\emph{B.3}] This step implements the Metropolis-Hastings label-switching moves detailed in Section \ref{sec:mixing}. These moves update $\bV^A$, $\bTheta^A$ and $\bZ$ jointly from their conditional distribution with $\bU$ integrated out. These moves are conditional upon the values of $\bV^A$ and $\bTheta^A$ sampled in steps \emph{B.1} and \emph{B.2}. The third label switching move is proposed and implemented for the Dirichlet Process only. 
  \item[\emph{B.4}] Conditioning on the updated values of $\bV^A$,$\bTheta^A$ and $\bZ$ from step \emph{B.3}, this step samples each $U_i$, $i=1,\ldots,n$, independently according to the full conditional distribution, $U_i\sim\mathrm{Unif}[0,\psi_{Z_i}]=\mathrm{Unif}[0,V_{Z_i}\prod_{l<Z_i}(1-V_l)]$, as detailed in \cite{W07}.
 \end{description}
 \item[\emph{Step C.}] To compute $U^{\star}$ is straightforward given the updated value of $\bU$ from step \emph{B.4}. The value of $Z^{\star}$ (and with it the set $A$) can only change from that computed in \emph{Step A} if the mixture component corresponding to the old $Z^{\star}$ was involved in a label switching move, and then only if the component it was switched with was empty. By design of the label switching moves (see Section \ref{sec:mixing}) this means that $Z^{\star}$ and $A$ can only get smaller, with the consequence that parameters corresponding to a small number of components may be updated twice per MCMC sweep (once in \emph{Step B} as part of the active components $A$, and once in \emph{Steps D} and \emph{E} as part of the updated potential components $P$). This has no ill-effects as long as the most recently updated parameter values are used at each subsequent step. 
\item[\emph{Step D.}] This is a joint update of $\alpha$, $\bV^P$ and $\bV^I$. The principle is to use the identity $p(\alpha,\bV^P,\bV^I|\cdot)=p(\alpha|\cdot)p(\bV^P,\bV^I|\alpha,\cdot)$. We proceed by first updating $\alpha\sim p(\alpha|\cdot)=\int p(\alpha,\bV^P,\bV^I|\cdot)\mathrm{d}\bV^P\mathrm{d}\bV^I$ (step \emph{D.1}) and then sampling $p(\bV^P,\bV^I|\alpha,\cdot)$ (step \emph{D.2}). To update $\bV^P$, we need to alternate Gibbs samples with checks to evaluate whether the component just updated is $C^{\star}$. In this way the set $P$ is determined on the fly. As mentioned in Section \ref{subsec:dpmmdefn}, no actual sampling is done for the inactive components in set $I$ as these would just be samples from the prior and have no impact on the likelihood or any other conditionals in the MCMC sweep. 

\begin{description}
\item[\emph{D.1}] Since $\bV^P$ and $\bV^I$ both correspond to empty mixture components, the only contribution to the joint posterior conditional is through the prior. This allows us to easily integrate out $\bV^P$ and $\bV^I$. Due to the conditional independence, the resulting posterior from which this step samples is $p(\alpha|\bV^A,\bZ)$. Typically this cannot be sampled directly, so we employ a Metropolis-within-Gibbs move to update $\alpha$, using an adaptive random-walk-Metropolis proposal on the log-scale.

If the prior for $\alpha$ is a Gamma distribution then it is alternatively possible to sample directly from the conditional with $\bV^A$ also marginalised (see \citeauthor{W07}, \citeyear{W07} and \citeauthor{EW95}, \citeyear{EW95} for details). We retain our version as any prior for $\alpha$ can be potentially used, even though only a Gamma prior is available in the code at the moment.

\item[\emph{D.2}] We begin by setting $C=Z^{\star}$. We then repeat the following two steps until the stop condition is reached. First, check if $\sum_{c=1}^C\psi_c > 1-U^{\star}$. Next, if the condition is met we set $C^{\star}=C$ and stop, otherwise we set $C=C+1$ and sample $V_C\sim\mathrm{Beta}(1-d,\alpha+dC)$. The Dirichlet process is a special case of the Pitman-Yor process for $d=0$. 
\end{description}
\item[\emph{Step E.}] This step updates the parameters $\bTheta^P$ and $\bTheta^I$ from the distribution $p(\bTheta^P,\bTheta^I|\cdot)$. The set $P$ is fully determined from step \emph{D.2}. The parameters correspond to empty mixture components, so updated values of $\bTheta^P$ are sampled directly from the prior. As with other inactive parameter $\bTheta^I$ play no part in this MCMC sweep and so are not updated.
\begin{description}
\item[\emph{E.1}] Taking advantage of the conditional independence structure, in this step we update $\bTheta^P$ by doing a Gibbs sample from the prior, such that $\Theta_c\sim p(\Theta_c|\Theta_0)$ for each $c\in P$. As $\Theta_c$ may be a vector of parameters, this may involve a number of Gibbs updates per component $c$. The full details depend upon the choice of $P_{\Theta_0}$. See Section \ref{sec:models} for examples.
\end{description}
\item[\emph{Step F.}] Here, the global (non-cluster-specific) likelihood parameters $\Lambda$ associated with $f$ are updated. There are only a finite number of such parameters so no special updates are needed.
\begin{description}
\item[\emph{F.1}] Sample $\Lambda\sim p(\Lambda|\bTheta^A,\bZ,\bD)$. Due to the conditional independence structure of the model the update only depends on the current value of the active likelihood parameters $\bTheta^A$, the allocations $\bZ$, and the data $\bD$. $\Lambda$ may contain multiple parameters, so this stage may contain many Gibbs and / or Metropolis-within-Gibbs steps. Full details will depend upon the choice of $f$ and $p(\Lambda)$, see Section \ref{sec:models} for examples.
\end{description}
\item[\emph{Step G.}] The final step of the algorithm is to update the parameter allocations $\bZ$, conditional on the newly updated values of the other parameters.
\begin{description}
\item[\emph{G.1}] We sample $\bZ\sim p(\bZ|\bV^A,\bV^P,\bTheta^A,\bTheta^P,\bU,\Lambda,\bD)$. Because of the independence of the individuals $i$, this is a series of Gibbs updates for each $i=1,2,\ldots,n$, sampling $Z_i\sim p(Z_i=c|\bV^A,\bV^P,\bTheta^A,\bTheta^P,U_i,\Lambda,D_i)$ where $D_i$ is the data for individual $i$. For each update, since the conditional $p(Z_i=c|\cdot)$ has no posterior mass for clusters $c$ where $U_i>\psi_c$, this update depends only on the parameters associated with the finite number of clusters in the sets $A$ and $P$, making the update a simple multinomial sample according to a finite vector of weights. The full details of the weights will depend upon the choice of $f$ and $P_{\Theta_0}$.
\end{description}   
\end{description}

\subsection{}
\label{appendix:contvarselect}
In the case of variable selection for continuous covariates, define the $J\times J$ matrix $\Gamma_c$ as
\[
   \Gamma_{c,i,j} = \begin{cases}
                     \gamma_{c,j} & i=j;\;\;\; j=1,2,\ldots,J\\
                     0 & \textrm{otherwise},
                    \end{cases}
\]
for the first variable selection method, and 
\[
   \Gamma_{c,i,j} = \begin{cases}
                     \zeta_j & i=j;\;\;\; j=1,2,\ldots,J\\
                     0 & \textrm{otherwise.}
                  \end{cases}
\]
for the second method as presented in Section \ref{section:varselect1}. Let $I_J$ denote the $J\times J$ identity matrix, $n_c$ be the number of individuals allocated to cluster $c$, $\bar{X}$ be as defined in Section \ref{subsubsec:contvarselect} and $\bar{X}_c=(\bar{X}_{c,1},\bar{X}_{c,2},\ldots,\bar{X}_{c,J})$ such that $\bar{X}_{c,j}=\sum_{i:Z_i=c}X_{i,j}/n_c$. The posterior conditional distributions for updating $\mu_c$ for $c\in A$ are then given by
\[
   \mu\sim\mathrm{Normal}(\tilde{\mu},\tilde{\Sigma})
\]
where 
\[
\tilde{\Sigma} = \left(\Sigma_0^{-1}+n_c\Gamma\Sigma_c^{-1}\Gamma\right)^{-1},   
\]
and 
\[
\tilde{\mu} = \tilde{\Sigma}\left[\Sigma_0^{-1}\mu_0 +n_c\Gamma\Sigma_c^{-1}(\bar{X}_c-(I_J-\Gamma)\bar{X})\right].   
\]

 \end{appendix}

\bibliography{dirichletSamplerPaper_2013}

\end{document}